\documentclass[10pt]{amsart}
\usepackage[dvips]{graphicx}
\usepackage{amsmath}
\usepackage{amsthm}
\usepackage{amssymb}
\usepackage{bm}
\usepackage{amstext}
\usepackage{psfrag}
\usepackage{amsfonts}

\newcommand{\p}{\partial}

\newcommand{\dd}{{\rm d}}

\newcommand{\bd}{\begin{definition}}                
\newcommand{\ed}{\end{definition}}                  
\newcommand{\bc}{\begin{corollary}}                 
\newcommand{\ec}{\end{corollary}}                   
\newcommand{\bl}{\begin{lemma}}                     
\newcommand{\el}{\end{lemma}}                       
\newcommand{\bp}{\begin{proposition}}            
\newcommand{\ep}{\end{proposition}}                
\newcommand{\bere}{\begin{remark}}                  
\newcommand{\ere}{\end{remark}}                     
\newtheorem{theorem}{Theorem}[section]
\newtheorem{corollary}[theorem]{Corollary}
\newtheorem{lemma}[theorem]{Lemma}
\newtheorem{proposition}[theorem]{Proposition}
\theoremstyle{definition}
\newtheorem{definition}[theorem]{Definition}
\theoremstyle{remark}
\newtheorem{remark}[theorem]{Remark}

\begin{document}


\title{ Classical aspects of lightlike dimensional reduction}

\author{E. Minguzzi }\thanks{Department of Applied Mathematics, Florence
 University, Via S. Marta 3, I-50139 Florence, Italy. {E-mail:
ettore.minguzzi@unifi.it}}


\maketitle

\begin{abstract}
\noindent Some  aspects of lightlike dimensional reduction in flat
spacetime are studied with emphasis to classical applications.
Among them the Galilean transformation of shadows induced by
inertial frame changes  is studied in detail by proving that, (i)
the shadow of an object has the same shape in every
orthogonal-to-light screen, (ii) if two shadows are simultaneous
in an orthogonal-to-light screen then they are simultaneous in any
such screen.  In particular, the Galilean group in 2+1 dimensions
is recognized as an exact symmetry of Nature which acts on the
shadows of the events instead that on the events themselves. The
group theoretical approach to lightlike dimensional reduction is
used to solve the reconstruction problem of a trajectory starting
from its acceleration history or from its projected (shadow)
trajectory. The possibility of obtaining a Galilean projected
physics starting from a Poincar\'e invariant physics is stressed
through the example of relativistic collisions. In particular, it
is shown that the projection of a relativistic collision between
massless particles gives a non-relativistic collision in which the
kinetic energy is conserved.
\end{abstract}

%


\newpage

\begin{flushright}
[\dots] {\em they see only their own shadows, or the shadows of
one another, which the fire throws on the opposite wall of the
cave} [\ldots]{\em . To them, I said, the truth would be literally
nothing but the shadows of the images.} \\
 Plato, The allegory of the cave,\\  Book VII of the
Republic
\end{flushright}


\section{Introduction}

In the famous book {\em The Republic} Plato, talking to one  of his
followers, Glaucon, introduced a powerful image, {\em The allegory
of the cave}, in order to explain how much the knowledge can make
the man free.

Plato imagined some ``observers" chained since their childhood and
constrained to look in front of them at a wall in the deep of a
cave. Light came behind them projecting their own shadows and the
shadows of other objects and people on the wall. According to
Plato the chained observers would identify themselves and the
other objects with their respective shadows, eventually loosing
their ability to perceive the third space dimension.

To the physical minded reader Plato's allegory represents the first
historical example of a dimensional reduction process. Indeed,
Plato's thought experiment represents a particular type of
dimensional reduction, that is, a {\em lightlike} dimensional
reduction. It differs from that considered much later by Kaluza and
Nordstr\"om \cite{oraifeartaigh00} (spacelike dimensional reduction)
or from that used in the hydrodynamic formalism of general
relativity or in the study of stationary metrics (timelike
dimensional reduction). In this work we shall study how the physics
laws on the full spacetime reduce to the quotient spacetime. A
general feature will be that relativistic physics projects to
non-relativistic physics. In Plato's allegory terms: the chained
observers would not be able to observe any relativistic effect,
independently of the speed reached by the shadows, that is, at any
degree of accuracy.
%

Our notations are as follows. Let $d \in \mathbb{N}$. The indexes
$i,j,k$, take the values $1,\ldots, d+1$,  the indexes $a,b,c$,
take the values $1,\ldots, d$. Vectors in $d$-dimensional vector
spaces are denoted in boldface, so that $v^{a}$ reads ${\bf v}$.
The Greek indexes $\alpha,\beta,\mu,\nu$, take the values
$0,1,\ldots, d+1$, and the indexes ${ A},{ B},{C}$, take the
values $0,1,\ldots, d$. For compactness the transpose of a vector
is denoted with the same letter. On Minkowski spacetime $M$ of
dimension $(d+1)+1$, we use coordinates $\{ x^{\mu}\}$,
$\mu=0,1,\ldots, d+1$,
 the spacelike convention $\eta_{00}=-1$, and units such that
$c=1$. In the list of commutation relations defining a Lie algebra
we shall omit the vanishing ones.

Mathematically the study of lightlike dimensional reductions began
in a 1929 work by Eisenhart \cite{eisenhart29} who  showed that
the trajectories of a Lagrangian system with $d$ degrees of
freedom
\begin{equation}
L(q^a\!,\,
\dot{q}^a\!,\,t)=\frac{1}{2}a_{ab}(q)\dot{q}^a\dot{q}^b+b_c(q)\dot{q}^c-U(q)
,
\end{equation}
may be obtained as the projection of geodesics of a
$d+2$-dimensional Lorentzian \footnote{The Lorentzianity of the
metric follows immediately by introducing the base of 1-forms
$\omega^{0}=\dd q^{0}$, $\omega^a=\dd q^a$, $\omega^{d+1}=\dd
q^{d+1}+(U+1/2)\dd q^{0}-b_c \dd q^c$ and by taking into account
the positive definiteness of $a_{bc}$. } spacetime of metric
\cite{eisenhart29,lichnerowicz55} \footnote{In our convention the
roles of $q^0$ and $q^{d+1}$ are inverted with respect to
\cite[Book II, Sect. 11]{lichnerowicz55} and there is also a
different choice of sign.}
\begin{eqnarray} \label{eis}
\dd s^2&=&a_{ab}\dd q^a\dd q^b +2 b_c\dd q^c \dd q^{0} -(2U+1)
(\dd q^{0})^2-2\dd q^{0}\dd q^{d+1} ,
\end{eqnarray}
where $q^{0}=t$ and $q^{d+1}$ is an auxiliary variable.

 In particular given a solution of the
dynamical system $q^{a}(t)$, set
\begin{equation} \label{dd}
q^{d+1}(t)=C- t+\int_0^t L(q^{a}(t),\dot{q}^{a}(t)) \dd t ,
\end{equation}
where $C$ is an arbitrary constant. The trajectory $\{q^{0},
q^{a}(q^0), q^{d+1}(q^0)\}$ can be regarded as an affine
parametrization of a
geodesic of the Eisenhart metric (\ref{eis})
with respect to a natural parameter \cite{lichnerowicz55},
$s=t=q^{0}$. Moreover, every solution of the Lagrangian system can
be regarded in this way. The Eisenhart metric takes its simplest
and most symmetric form in the case of a free classical particle
$a_{bc}=\delta_{bc}$, $b_c=0$, $U=0$. Remarkably in this case the
Eisenhart metric becomes the Minkowski metric as can be seen
introducing alternative coordinates ${x}^{i}=q^i$,
$x^0=q^0+q^{d+1}$.


This result was the first signal of an interesting correspondence
between Poincar\'e invariance and Galilei invariance in a lightlike
reduced spacetime with one dimension less. Nevertheless, it took
half a century to fully realize the correspondence. Some progress
was made after the impulse of Dirac's work \cite{dirac49} that led
to the so called {\em front wave}  (or {\em light front} or {\em
infinite momentum frame}) dynamics
\cite{bouchiat71,leutwyler78,harindranath96,kutach04}. Later, a work
by Bargmann \cite{bargmann54} inspired a series of works where this
correspondence was fully appreciated and led to  the  {\em Bargmann
structure} approach to the  Newton-Cartan theory
\cite{kunzle72,duval85,duval91,bernal03b,ehlers97}. The algebraic
conditions at its foundations were slightly weakened in subsequent
works \cite{julia95}. In both lines of research the authors were
involved in quantum mechanical problems. In the former case the
authors studied an alternative formulation of quantum field theory
by using null Cauchy surfaces. In the latter case the authors
considered the $(d+1)+1$ spacetime mainly as a tool for expressing
in a simple way the symmetries of Galilei invariant quantum
mechanical systems in $d+1$ spacetime dimensions.

In this work we focus on  classical applications of lightlike
dimensional reduction. Indeed, we feel that already at the classical
level this structure may have interesting applications. For
instance, the correspondence between Poincar\'e invariance and
Galilei invariance in a lightlike reduced spacetime  may be grasped
particularly well in its simplest application: the study of the
transformation properties of shadows under inertial frame changes.
In order to keep the work at a reasonable size we shall limit
ourselves to the flat spacetime case as it has some peculiarities
which are worth studying in their own right. Indeed, in Minkowski
spacetime the Poincar\'e group is left unbroken and the group
theoretical approach becomes particularly advantageous. On the
contrary, in the curved spacetime case differential geometric tools
should be preferred, and new concepts such as the mentioned Bargmann
structures or Eisenhart's spacetime should be introduced.

This work is organized as follows.
\begin{itemize}
\item In section \ref{unov} we  model our problem
by considering a distant source of light and some sets of inertial
observers suitably oriented with respect to the direction of light
(subsection \ref{uno}). Then the groups acting transitively on
each set are studied (subsection \ref{due}), and it is shown that
suitably accelerating and rotating observers can have comoving
inertial frames which keep staying, time by time, in the same
original set (subsection \ref{due}). In subsection \ref{quat} it
is shown that a subgroup of the Poincar\'e group is in fact, a
central extension of the Galilei group. Thus, although the Galilei
group does not act on events, its action is well defined on the
quotient space $Q$ of their shadows. The application to the
transformation properties of shadows is studied in detail
(subsections \ref{cinq} and \ref{sei}), and several conclusions
are finally drawn (subsection \ref{sett}).
\item Section \ref{ott} is devoted to various reconstruction
problems. First, the problem of reconstructing the trajectory
starting from the decomposition of the proper acceleration in
acceleration along the direction of light and orthogonal to it is
considered. A new particular (lightlike) parallel transport
related to the presence of a preferential lightlike vector is
introduced by means of group theoretical methods, and it is shown
to give rise to an absolute transverse orientation (subsection
\ref{ott2}). The new parallel transport is completely integrated
and the relevance of this solution for the problem of autonomous
spacetime navigation is emphasized (subsection \ref{ott3}). In
subsection \ref{ott4} the problem of reconstructing a trajectory
starting from its shadow is considered, and the relation with the
classical action is pointed out. In subsection \ref{ott5} a
formula for the time dilation given the acceleration history of a
non-inertial observer is obtained. It is the first formula of this
kind developed in three or more space dimensions, the analogous
problem for a decomposition of the acceleration with respect a
Fermi-Walker triad being still open. In the last subsection
\ref{ott6} we point out that the knowledge, by an inertial
observer, of the sky position and longitudinal frequency of a
signal emitted by a source does not allow to deduce the behavior
of the distance. A position drift may always occur which is
related to the classical action.
\item Section \ref{fp} is devoted to an example in which the
projection of a Poincar\'e symmetry into a Galilei symmetry becomes
particularly clear, that is, that of a relativistic collision which
projects into a non-relativistic collision. The reduced Lagrangian
is calculated showing that it is of classical type and the kinetic
and internal energies are identified. It is shown that any shadow
worldline has an associated (inertial reference frame) invariant,
i.e. the {\em shadow mass}, which plays the role of a
non-relativistic mass. The conservation of shadow mass, momentum and
energy is shown to follow from the conservation of relativistic
momentum in the full spacetime. Although the kinetic energy is not
necessarily conserved, it is shown that it is conserved in the
projection of a relativistic collisions between massless particles.
Finally, the inverse problem of finding the relativistic collision
from which a non-relativistic collision comes from, is solved.
\item In section \ref{conc} we give some conclusions.
\item In Appendix \ref{lpt} a differential geometric formulation
of the lightlike parallel transport is given.
\item In Appendix \ref{pol} the formal analogy between the problem of
transforming shadows between different inertial frames, and the
problem of transforming the photon polarization vector is pointed
out. Some ambiguities regarding the role of the transverse
Galilean boosts generators in this context are clarified.

\end{itemize}

\section{Shadows and group aspects of lightlike dimensional
reduction} \label{unov}

In this section
the reader is introduced to the geometry of lightlike dimensional
reduction through the study of shadows and their transformation
properties.
However, we stress that the results of this section are not
confined to the application to shadows, but are rather general
properties of the geometry of  lightlike dimensional reduction.

\subsection{Subsets of inertial observers} \label{uno}
let $M$ be Minkowski spacetime. By {\em observer} we mean a timelike
worldline $\gamma(\tau)$ parametrized with respect to proper time
and an orthonormal tetrad field $\{e_0=\p_{\tau}, e_i\}$ over it. If
the tetrad is parallely transported the observer is inertial. An
orthonormal tetrad at $m \in M$, $\{e_0, e_i\}$ with $e_0$ timelike
will  be said to be an {\em inertial frame} since  canonical
coordinates $\{x^{\mu}\}$ with origin $m$ can be introduced such
that $e_{\mu}=\p/\p x^{\mu}$, $\dd s^2=\eta_{\mu \nu}\dd x^{\mu} \dd
x^{\nu}$. Sometimes an inertial frame will be denoted with the
letter $K$. Thus every observer is a sequence of inertial frames
$K(\tau)$. The inertial frames of an inertial observer are related
by time translations.

 Consider a point like source of monochromatic light $\Sigma$
placed very far from a spacetime region $\mathcal{U} \subset M$ of
Minkowski spacetime. For an inertial observer inside $\mathcal{U}$
the light coming from the source is described by a congruence of
parallel null geodesics, the photons coming from the source having a
well defined momentum $p^{\mu}$. The inertial observers inside
$\mathcal{U}$, without changing their covariant velocity, can orient
their own space axes $\{ e_i \}$, so that the photons travel in
direction ${e}_{d+1}$, $p^{\mu}=\omega(1,0,\ldots,0,1)$ with a
constant $\omega$ possibly depending on the observer. We are
focusing here on the subset $S(\hat{p})$ of the inertial frames for
which the null momentum $p^{\mu}$ has the same spatial components up
to a multiplicative constant. The subset $S(\hat{p})$ splits into
subsets $S_{\omega}(\hat{p})$ dependent on the value of the
frequency $\omega$ measured by the observer, $p^i=\omega \hat{p}^i$.


Let $L({e}_{d+1})\subset SO(1,d+1)$ be the little group of the
null vector $n^{\mu}=(1,0,\ldots,1)$ (by $SO(1,d+1)$ we mean the
connected component of the Lorentz group containing the identity),
that is, the subgroup of $SO(1,d+1)$ which leaves $n^{\mu}$
invariant, and analogously let $IL({ e}_{d+1}) \subset ISO(1,d+1)$
be the little group with the translations included, $L({e}_{d+1})
\subset IL({e}_{d+1})$. Moreover, let $G({e}_{d+1})$ be the group
obtained from $L({e}_{d+1})$ by including the boosts in the
direction ${e}_{d+1}$, i.e. the group of those Lorentz
transformations that send $n^{\mu}$ to a vector proportional to
it, and let $IG({e}_{d+1})$ be the group obtained by including the
boosts in direction ${e}_{d+1}$ to $IL({e}_{d+1})$, or the
translations to $G({e}_{d+1})$. It is known \cite{weinberg95} that
the group $IL({e}_{d+1})$ acts freely and transitively on
$S_{\omega}({e}_{d+1})$ and that the group $IG({e}_{d+1})$ acts
freely and transitively on $S({e}_{d+1})$. Under changes of $n$
the groups remain the same up to isomorphisms.

The group $IG({e}_{d+1})$ represents Poincar\'e transformations
between inertial frames for which the light from $\Sigma$ comes from
the same direction $-{e}_{d+1}$. The group $IL({e}_{d+1})$
represents Poincar\'e transformations between inertial frames for
which the light from $\Sigma$ comes from the same direction
$-{e}_{d+1}$ with the same frequency. The groups $G({e}_{d+1})$ and
$L({e}_{d+1})$ add the condition that the related inertial frames
share the same origin of coordinates.

We assume that the inertial observers set up semitransparent screens
perpendicular to the direction of light, at their respective
coordinate $x^{d+1}=0$ in such a way that light comes from the
region $x^{d+1}<0$. In this way the shadow of an object projects on
their respective screens leaving a dark image on each one. One
should be careful because, although the screens of different
observers are perpendicular to the same direction of light, the
screens are in no sense parallel due to the aberration of light
under boosts. Our aim is to show that the different images of the
same shadow are related  between the screens of observers in
$S_{\omega}({e}_{d+1})$ by a  Galilean transformation and between
the screens of observers in $S({e}_{d+1})$ by a Galilean
transformation plus a time scaling.

\begin{figure}
\centering \psfrag{A}{$\Sigma$} \psfrag{B}{$v$} \psfrag{C}{$x^1$}
\psfrag{F}{$x^2$} \psfrag{E}{$x^{d+1}$} \psfrag{D}{${\bf v}$}
\includegraphics[width=8cm]{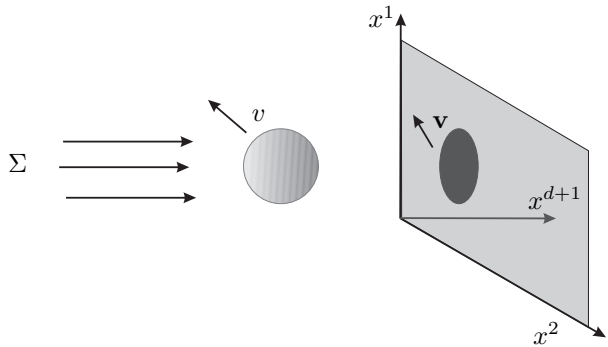}
\caption{Shadow of an object on the plane perpendicular to the last
axis. The worldlines of the points at rest with the plane span on
spacetime the hyperplane $x^{d+1}=0$. Assume the plane is a
semitransparent screen so that the shadow has different images, one
for each orthogonal-to-light screen. How does the shadow image
change under change of orthogonal-to-light screen? }
\end{figure}

This problem is somewhat related to that of finding how the
appearance of the night sky transforms under Lorentz transformations
\cite{penrose59,rindler84,naber92}. Whereas there one focuses on
light reaching an event from all directions, here we focus on light
of a given direction projecting on planes.

\subsection{The little group of the massless particle} \label{due}
Let us briefly recall how to construct $L({e}_{d+1})\,$
\cite{weinberg95}. Let
\begin{equation} \label{poi2}
{x'}^{\mu}=L^{\mu}_{\ \nu} x^{\nu} -b^{\mu} ,
\end{equation}
be a transformation in $IL({e}_{d+1})$. The invariance  of the
(d+2)-momentum $\omega n^{\mu}$ reads $L^{\mu}_{\ \nu} n^{\nu}=
n^{\mu}$. Let $t^{\mu}=(1,0,\ldots,0)$, the vector
$(Lt)^{\mu}=L^{\mu}_{\ \nu} t^{\nu}$ is a unit vector that
satisfies $n_{\mu} (Lt)^{\mu}=(Ln) \cdot (Lt)=n\cdot t=-1$. As a
consequence $Lt$ can be written
\begin{equation}
(Lt)^{\mu}=(1+\zeta,-\alpha^1, \ldots, -\alpha^{d}, \zeta) ,
\end{equation}
for a suitable $\alpha^a$ and
$\zeta=\boldsymbol{\alpha}\!\cdot\!\boldsymbol{\alpha} /{2}$. The
Lorentz transformation
\begin{equation}\label{lk2}
S(\boldsymbol{\alpha})^{\mu}_{\ \nu}\!=\!\begin{pmatrix}
1+\zeta & -\boldsymbol{\alpha} & -\zeta \\
 -\boldsymbol{\alpha}  & I  &  \boldsymbol{\alpha}  \\
\zeta & -\boldsymbol{\alpha}  & 1-\zeta
\end{pmatrix},
\end{equation}
 where $I^{a}_{\ b}$ is the identity matrix, leaves $n^{\nu}$
invariant and sends $t$ to $Lt$. Thus the Lorentz transformation
$L^{-1}S$ leaving invariant both $t$ and $n$ (and hence
${e}^{\mu}_{d+1}=n^{\mu}-t^{\mu}$) must be an element of $SO(d)$,
that is a rotation  $(R^{-1})^{\mu}_{\ \nu}$ that leaves
${e}_{d+1}$ invariant. The matrix $R^{\mu}_{\ \nu}$ takes the
form, with obvious notation,
\begin{equation} \label{rot}
R^{\mu}_{\ \nu}\!=\!\begin{pmatrix}
1 & {\bf 0} & 0 \\
 {\bf 0} & \textrm{R} &  {\bf 0} \\
0 & {\bf 0} & 1
\end{pmatrix},
\end{equation}
where $\textrm{R}^{a}_{\ b}=\textrm{R}_{ab} \in SO(d)$. The
generic element $L^{\mu}_{\ \nu}$ of the little group
$L({e}_{d+1})$ takes the form $L^{\mu}_{\ \nu}=S^{\mu}_{\ \beta}
R^{\beta}_{\ \nu}$ or
\begin{equation} \label{gf}
L^{\mu}_{\ \nu}\!=\!\begin{pmatrix}
1+\zeta & -{\alpha}^{c} \textrm{R}_{c b} & -\zeta \\
 -{\alpha}^{a}   & \textrm{R}^{a}_{\ b}  &  \alpha^a  \\
\zeta & -\alpha^{c}\textrm{R}_{c b}  & 1-\zeta
\end{pmatrix}.
\end{equation}
Note that the matrices of type (\ref{lk2}) form an Abelian
subgroup of dimension $d$, the so called `translations' of the
little group. From  the expression (\ref{gf}) we can recover the
Lie algebra $\mathcal{L}({e}_{d+1})$ of $L({e}_{d+1})$. The
infinitesimal transformation has the form ($J^{ab}=J_{ab}$)
\begin{equation}
L\simeq I+\frac{1}{2}\Omega_{a b}J^{a b}+ \alpha^{a} W_{a}
\end{equation}
where $W_a$ are the generators of `translations' and (here and
throughout the work we omit vanishing commutation relations such
as $[W_a,W_b]=0$ while reporting the commutation relations of an
entire Lie algebra)
\begin{eqnarray}
{}[J_{ab},J_{cd}] &=&\delta_{ad}J_{b c} +\delta_{b d} J_{a
d}-\delta_{a c} J_{b d} -\delta_{b d} J_{a c} , \label{l1}\\
{}[W_{a}, J_{b c}] &=& \delta_{a b} W_{c}-\delta_{a c} W_{b} ,
\label{l2}
\end{eqnarray}
thus $L({e}_{d+1})$ is isomorphic to the group $ISO(d)$. From
(\ref{gf}) one can also find the expression of these generators in
terms of the generators of the Lorentz group
\begin{eqnarray}
{}[J^{\alpha \beta}\!, J^{\gamma \delta}] \!\!&=&\! \!\eta^{\alpha
\delta} J^{\beta \gamma}\!+\!\eta^{\beta \gamma} J^{\alpha
\delta}\!\!-\!\eta^{\alpha \gamma}J^{\beta \delta}\!\!-\!
\eta^{\beta \delta} J^{\alpha \gamma}.
\end{eqnarray}
Indeed, the usual matrix representation of the group
${x'}^{\mu}=\Lambda^{\mu}_{\ \nu} x^{\nu}$ induces the Lie algebra
representation $(J^{\alpha\beta})^{\mu}_{\ \nu}=\eta^{\alpha \mu}
\delta^{\beta}_{\nu}-\eta^{\beta\mu} \delta^{\alpha}_{\nu}$.
From the infinitesimal version of Eq. (\ref{gf}) we obtain
\[
W_{a}=J^{0 \, a}-J^{d+\!1 \, a} ,
\]
and, as the notation suggests, $J_{ a b}$ is nothing but $J_{\mu
\nu}$ with the indexes restricted to the values $1, \ldots d$.
Moreover, as expected, $(J_{a b})^{\mu}_{\
\nu}n^{\nu}=(W_a)^{\mu}_{\ \nu}n^{\nu}=0$.

\subsection{Poincar\'e group, inhomogeneous little group and
non-inertial observers}\label{obs}

The infinitesimal  Poincar\'e transformation
${x'}^{\mu}=\Lambda^{\mu}_{\ \nu} x^{\nu}-b^{\mu}$  in a $(d+1)+1$
spacetime can be written
\begin{equation}
I+\frac{1}{2}\Omega_{\alpha \beta}J^{\alpha \beta}- b_{\gamma}
P^{\gamma} ,
\end{equation}
where the generators $J^{\alpha \beta}$, $P^\gamma$, satisfy the
Lie algebra
\begin{eqnarray}
{}[J^{\alpha \beta}, J^{\gamma \delta}] \!&=& \!\eta^{\alpha
\delta} J^{\beta \gamma}\!+\eta^{\beta \gamma} J^{\alpha
\delta}\!-\eta^{\alpha \gamma}J^{\beta \delta}\!- \eta^{\beta
\delta}
J^{\alpha \gamma},\\
{}[P^{\alpha} , J^{\beta \gamma}] \! &=& \! \eta^{\alpha
\beta}P^{\gamma} - \eta^{\alpha \gamma} P^{\beta} \label{jhg} .
\end{eqnarray}
We shall write $H=P^0$.


Any non-inertial observer passes through a sequence of inertial
frames $K(\tau)$.  The coordinate transformation between $K(\tau)$
and $K(\tau +\dd \tau)$ is an infinitesimal Poincar\'e
transformation $I+\frac{1}{2}\tilde\omega_{\alpha \beta}J^{\alpha
\beta}\dd \tau+H\dd \tau$ where the space translations do not enter.
Indeed, the presence of space translations would imply a violation
of causality. Here the motion is regarded as a sequence of
infinitesimal time translations, boosts and rotations. The infinite
product of those infinitesimal transformations may generate space
translations that, however, are not present in the infinitesimal
transformation.

Let ${x}_{K(\tau)}^{\mu}=\Lambda^{\mu}_{\ \nu}(\tau)
[x^{\nu}_{K(0)}-x^{\mu}(\tau)]$ be the Poincar\'e transformation
from $K(0)$ to $K(\tau)$. Here $x^{\mu}_{K(\tau)}$ are the
coordinates of $K(\tau)$, and $x^{\mu}(\tau)$ represents the
origin of $K(\tau)$ with respect to $K(0)$. This  general
transformation arises from an infinite product of  the said
infinitesimal transformations.
 The covariant velocity of $K(\tau)$ in its own
coordinates is $\delta^{\mu}_{0}$ hence the covariant velocity of
$K(\tau)$ in $K(0)$'s coordinates is
$u^{\mu}(\tau)=(\Lambda^{-1})^{\mu}_{\ 0}$. We have
\begin{eqnarray}
 \frac{\dd}{\dd
\tau}\Lambda&=&\frac{1}{2}\tilde\omega_{\alpha
\beta}(\tau)J^{\alpha
\beta}\Lambda, \\
x^{\mu}(\tau)&=&\int_0^{\tau}\dd \tau'(\Lambda^{-1}(\tau'))^\mu_{\
0} .
\end{eqnarray}
The motion of the non-inertial observer can be recovered from the
knowledge of $\tilde{\omega}_{\alpha \beta}(\tau)$. Its physical
meaning is the following. The acceleration of the non-inertial
observer in $K(0)$'s coordinates is $\frac{\dd}{\dd
\tau}(\Lambda^{-1}(\tau'))^\mu_{\ 0}$ hence the acceleration of the
non-inertial observer in its own coordinates is
\[
\tilde{a}^{\mu}=\Lambda^{\mu}_{\ \nu} \frac{\dd}{\dd
\tau}(\Lambda^{-1})^\nu_{\ 0}=-\tilde\omega^{\mu}_{\ \, 0} .
\]
The tilde reminds us that $\tilde{a}^\mu$ is not only the covariant
acceleration $a^\mu$, which makes sense in arbitrary frames, but the
covariant acceleration in  the coordinates of the accelerating frame
(hence $\tilde{a}^{0}=0$). The components of the acceleration in the
non-inertial observer's axes are $\tilde\omega^{\mu
0}=\tilde\omega_{0\mu}$. Analogously
$\tilde\omega^{ij}=\tilde\omega_{ij}$ represents the tensorial
angular velocity of the non-inertial frame in its own coordinates.

Now, let us consider a non-inertial observer such that $K(\tau)$
passes through inertial frames belonging to $S_{\omega}(e_{d+1})$.
Let ${x}_{K(\tau)}^{\mu}=L^{\mu}_{\ \nu}(\tau)
[x^{\nu}_{K(0)}-x^{\mu}(\tau)]$ be the inhomogeneous little group
transformation from $K(0)$ to $K(\tau)$. This time we have that
$L(\tau+\dd \tau)=(I+\frac{1}{2}\tilde{\omega}_{a b}\dd \tau J^{a
b}+ \tilde\alpha^{a} \dd \tau W_{a})L$. Using the expression for
$W_a$,  we obtain the constraints
\begin{eqnarray}
\tilde{a}^{d+1}&=&\tilde\omega^{d+\!1\, 0}=0,\\
\tilde{a}^{a} &=&\tilde\omega_{0a}=-\tilde\omega_{d+\!1 \,
a}=\tilde\alpha_a .
\end{eqnarray}
In other words the non-inertial observer may keep staying time by
time into $S_{\omega}(e_{d+1})$ despite its non-inertial motion
provided (i) its acceleration along the direction $e_{d+1}$ vanishes
and (ii) (this interpretation holds only for $d =2$ in which case an
angular velocity vector can be defined), the projection of the
angular velocity vector on the plane perpendicular to $e_{d+1}$ and
the acceleration are perpendicular to each other and of the same
magnitude, moreover, their vector product  equals $e_{d+1}$ up to a
non-negative factor.

Due to the different dimensionality of acceleration and angular
velocity we find, restoring $c$, that the projected angular velocity
is indeed very small. It is required in order to correct for the
aberration of light due to the acceleration which would change the
night sky position of the source $\Sigma$ from the direction
$-e_{d+1}$.

\subsection{The Galilei quotient group} \label{quat} Let
\begin{eqnarray}
\mathcal{M}&=& H-P^{d+1} ,
\end{eqnarray}
in such a way that the infinitesimal transformation in
$\mathcal{IL}(e_{d+1})$ reads
\[
I+\frac{1}{2}\Omega_{a b}J^{a b}+ \alpha^{a} W_{a}- b^{b}P_{b}
+(b^{0}\!-b^{{d+1}})H+b^{d+1}\mathcal{M} . \]

The Lie algebra $\mathcal{IL}(e_{d+1})$ in terms of the generators
$J_{a b}$, $W_a$, $P_a$, $H$ and $\mathcal{M}$ is obtained by
adding to Eqs. (\ref{l1})-(\ref{l2})  the commutations relations
that follow from Eq.~(\ref{jhg})
\begin{eqnarray}
{}[P_a,J_{b c}]&=&\delta_{a b} P_{c}-\delta_{a c} P_{b}, \label{l3} \\
{}[W_a, H]&=& P_a, \label{l4}\\
{}[W_a, P_b]&=&  \delta_{a b}\mathcal{M}. \label{l5}
\end{eqnarray}
The generators $J_{ab}$, $W_a$, $P_a$ and $H$ would form a Galilei
subalgebra $\mathcal{G}al(d)$ given by the only non-vanishing
commutation relations (\ref{l1}), (\ref{l2}), (\ref{l3})  and
(\ref{l4}) if the right hand side of (\ref{l5}) would vanish. Since
$\mathcal{M}$ commutes with the other generators, $\mathcal{IL}$ is
a central extension of the Galilei algebra $\mathcal{G}al(d)$ in
$d+1$ spacetime
\begin{equation}
0 \to \{\mathcal{M}\} \to \mathcal{IL}(e_{d+1}) \to
\mathcal{G}al(d) \to 0,
\end{equation}
where $\{\mathcal{M}\}$ is the 1-dimensional Lie algebra spanned
by $\mathcal{M}$.

The  inequivalent central extensions of the Galilei algebra have
been classified by Bargmann \cite{bargmann54} for the case $d \ge 3$
and by Bose \cite{bose95} for the case $d=2$. The vector space of
inequivalent  extensions is one dimensional in the former case and
three dimensional in the latter case. Thus, the central extension
considered here is the only one available for $d \ge 3$, and in
particular it is the one that makes sense physically. Indeed, in
quantum mechanics the unitary ray representations of a group $G$ are
induced by the unitary representations of the central extensions of
the universal covering of $G$, $G^*$. Bargmann has shown that the
projective unitary representation of the Galilei group that makes
sense physically  is the one obtained from the central extension
considered above. The operator $\mathcal{M}$, in that quantum
mechanical context, is known as the mass operator. Its presence
implies { the mass superselection} rule which forbids the
superpositions of states with different mass
\cite{levyleblond63,weinberg95}. We shall see below that it has the
same meaning in our classical context with the difference that by
{\em mass} here we shall mean the one that makes sense in a suitable
$d+1$ Galilean spacetime $Q$. In the application to shadows, for
instance, it will be called the {\em shadow mass},  to  distinguish
it from the mass of the particle that projects the shadow.

Let $N$ be the 1-dimensional normal subgroup generated by
$\mathcal{M}$, then $IL(e_{d+1})/N \sim {G}al(d)$ has the correct
Lie algebra. The central extension at the group level is
\begin{equation}
1 \to N (\sim T_1) \to IL(e_{d+1}) \to {G}al(d) \to 1.
\end{equation}
The group $Gal(d)$ does not act on events of coordinates $x^{\mu}$
but rather on the worldlines of the form $Nx$. Since
$\mathcal{M}=H-P^{d+1}=-P^{\gamma}n_{\gamma}$, the effect of an
element of $N$ on $x^{\mu}$ is that of adding a vector
proportional to $n^{\mu}$, i.e. the elements $Nx$ are the light
rays emitted from $\Sigma$. The reduced spacetime $Q$ is the space
of null geodesics of $M$ of direction $n^{\mu}$. The dimensional
reduction of $M$ onto $Q$ along a lightlike direction $n$ sends
the little group $IL(e_{d+1})$ into the Galilei group. Thus, at
least for the case $d=2$ we can regard the Galilean symmetry in
2+1 spacetime as an exact symmetry of Nature rather than as an
approximate symmetry for small velocities. In order to reveal this
symmetry one has to focus on the shadows of the objects in place
of the material objects themselves. On the contrary, the Galilean
symmetry of the full $3+1$ spacetime remains an approximate
symmetry valid only for small velocities.

\subsection{The transformation of shadows} \label{cinq}

\begin{figure}
\centering \psfrag{A}{$n$} \psfrag{B}{$t$} \psfrag{C}{$x^0$}
\psfrag{F}{$x^a$} \psfrag{E}{$x^{d+1}$} \psfrag{G}{$x$}
\includegraphics[width=6cm,height=6cm]{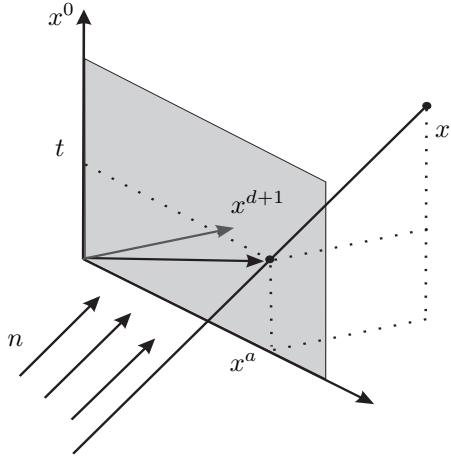}
\caption{The decomposition of the event of coordinates $x^{\mu}$
into transversal (shadow) position ${\bf x}$, hitting time $t$,
and longitudinal coordinate $x^{d+1}$. The interpretation of ${\bf
x}$ as the position of the shadow makes sense only if $x^{d+1}\le
0$. } \label{feg2}
\end{figure}

Chosen an inertial frame in $S_{\omega}({e}_{d+1})$ of coordinates
 $\{x^\mu\}$ every event $x^{\mu}$ admits the unique decomposition
 ($t=x^{0}-x^{d+1}$)
\begin{equation} \label{dec}
x^{\mu}=\begin{pmatrix} x^{d+1}+t\\
{\bf x} \\
x^{d+1}
\end{pmatrix} =
x^{d+1} n^{\mu} + \begin{pmatrix}
t\\
{\bf x} \\
0
\end{pmatrix} .
\end{equation}
The parameter $t$ represents the time at which the light beam from
$\Sigma$ passing through $x^{\mu}$ hits the surface (screen)
$x^{d+1}=0$. The vector ${\bf x}$ gives the hitting point on the
screen (see figure \ref{feg2}). By shadow of event $x$ we mean the
null geodesic of direction $n^{\mu}$ passing through $x$ or its
representative $(t,{\bf x})$. Note that physically the  shadow
exists only if the object of worldline $x^{\mu}(\tau)$ that
projects the shadow passes between the source and the screen,
$x^{d+1}(\tau)<0$. We shall not impose this condition because it
is not restrictive (the frame origin can be translated) and
because we are more interested on the mathematical definition of
shadow.

The shadow of a ($C^1$) particle worldline $x^{\mu}(\lambda)$ is
that shadow worldline ${\bf x}(t)= {\bf x}(\lambda(t))$ composed
of the shadows of the events of the original worldline. We shall
assume that $x^{\mu}(\lambda)$ is causal, future pointing and that
$\dd x^{\mu}/\dd \lambda$ is not proportional to $n^{\mu}$, that
is $\frac{\dd x^{\mu}}{\dd \lambda}n_{\mu}<0 \Rightarrow \dd t/\dd
\lambda>0$, then ${\bf x}(t)$ is future pointing in the sense that
$t(\lambda)$ is increasing. In this work we shall consider massive
and massless particles on $M$. One should be careful with the
massless particles having a four momentum proportional to
$n^{\mu}$, since they have a special role. They do not project on
worldlines ${\bf x}(t)$ but rather on isolated events $(t, {\bf
x})$. Thus such particles represent the events of our reduced
Galilean spacetime $Q$ rather than the particles moving on it.

Since the light ray passing though $x$ is determined by the
parameters $t$ and ${\bf x}$ one would expect that under a
transformation in $IL({e}_{d+1})$ these parameters transform
according to the Galilei quotient group. Indeed, this is the case.
If ${x'}$ is related to $x$ by Eq. (\ref{poi2}) with $L^{\mu}_{\
\nu}$ given by (\ref{gf}) then
\begin{equation}
{x'}^{\mu}=\begin{pmatrix} {x'}^{d+1}+t'\\
{\bf x}' \\
{x'}^{d+1}
\end{pmatrix} =
{x'}^{d+1} n^{\mu} + \begin{pmatrix}
t'\\
{\bf x}' \\
0
\end{pmatrix}
\end{equation}
with
\begin{equation} \label{r1}
{x'}^{d+1}=x^{d+1}+t \zeta-\alpha^a \textrm{R}_{ab} x^{b}-b^{d+1},
\end{equation}
and (Galilei transformation)
\begin{eqnarray}
t'&=&t-(b^0-b^{d+1}) ,\label{r2} \\
{x'}^b&=&\textrm{R}^{b}_{\ c} x^{c}-t \alpha^b-b^b . \label{r3}
\end{eqnarray}
The transformation (\ref{r1})-(\ref{r3}) makes it clear that the
generators $W_a$ generate Galilean boosts on the $d+1$ quotient
spacetime. Their commutativity expresses the commutativity of
Galilean boosts.

In the coordinates $x^{d+1}$, $t$ and $x^a$ the Minkowski metric
reads
\begin{equation}\label{mink}
\dd s^2=-\dd t^2-2\dd t \, \dd x^{d+1}+\dd x^a \dd x_a .
\end{equation}
It is  invariant  under the transformation (\ref{r1})-(\ref{r3}).

\subsection{The inclusion of boosts of direction $e_{d+1}$}
\label{sei} In the previous sections we studied the group
$IL(e_{d+1})$ that acts freely and transitively on
$S_{\omega}(e_{d+1})$. The boosts in direction $e_{d+1}$ do not
belong to this group since they would change the observed frequency
$\omega$. Thus the inertial frames belonging to
$S_{\omega}(e_{d+1})$, not only are suitably oriented but also their
covariant velocity satisfies a constraint and is not completely
general. The situation is quite different with the set $S(e_{d+1})$:
up to a reorientation of the axes every inertial frame belongs to
this set. It is important to generalize our conclusion to the group
$IG(e_{d+1})$ that acts freely and transitively on $S(e_{d+1})$ as
the results would hold for any observer as long as we identify an
observer with its worldline.

The subgroup $G(e_{d+1})$  of the Lorentz group sends the null
vector $n^{\mu}=(1,0,0,1)$  to a vector proportional to
it.\footnote{The group $G(e_{d+1})$ sends the null plane $n^\mu
x_\mu=0$ into itself. This fact explains the close relation between
this work and studies on  {\em front wave dynamics}.} Let $G^\mu_{\
\nu} \in G(e_{d+1})$, we have $G^\mu_{\ \nu} n^{\nu}=e^{-r(G)}
\,n^{\mu}$ for a suitable constant $r(G)$ (by Lorentz group we mean
the connected component which contains the identity, thus, since it
sends the forward light cone into the forward light cone, the
proportionality constant is positive). The ratio between the
frequencies in the two frames is $\omega'/\omega=e^{-r}$. Note that
the boost
\begin{equation} \label{boo}
B(r, e_{d+1})=
\begin{pmatrix}
  \cosh r &  {\bf 0} & -\sinh r \\

  {\bf 0} & \delta^{a}_{\ b}   & {\bf 0}\\
  -\sinh r & {\bf 0} & \cosh r
\end{pmatrix}.
\end{equation}
satisfies $B^{\mu}_{\ \nu}(r, e_{d+1}) n^{\nu}=e^{-r} n^{\mu}$.
Thus ${B^{-1}}^{\mu}_{\ \nu} G^{\nu}_{\ \alpha}n^{\mu}=n^{\mu}$,
that is $L^{\mu}_{\ \alpha}={B^{-1}}^{\mu}_{\ \nu} G^{\nu}_{\
\alpha}$ belongs to the little group $L(e_{d+1})$. The parametric
form of $G^{\mu}_{\ \nu}={B}^{\mu}_{\ \beta}L^{\beta}_{\ \alpha}$
is
\begin{equation} \label{gma}
G^{\mu}_{\ \nu}\!=\!\begin{pmatrix}
\cosh r+e^{-r}\zeta & -e^{-r}{\alpha}^{c} \textrm{R}_{c b} & -\sinh r - e^{-r} \zeta \\
 -{\alpha}^{a}   & \textrm{R}^{a}_{\ b}  &  \alpha^a  \\
-\sinh r +e^{-r} \zeta & -e^{-r}\alpha^{c}\textrm{R}_{c b}  &
\cosh r-e^{-r}\zeta
\end{pmatrix}.
\end{equation}
Using the decomposition (\ref{dec}) the generic $IG(e_{d+1})$
transformation, ${x'}^{\mu}=G^{\mu}_{\ \nu}x^{\nu}-b^{\mu}$, takes
the form
\begin{eqnarray}
{x'}^{d+1}\!\!\!\!&=&\!e^{-r}x^{d+1}+t (e^{-r}\!\zeta\!\!-\!\sinh
r)-e^{-r} \!\alpha^a
\textrm{R}_{ab} x^{b}-b^{d+1}, \nonumber \\
t'&=&e^{r} t-(b^0-b^{d+1}) , \label{d1}\\
{x'}^b&=&\textrm{R}^{b}_{\ c} x^{c}-t \alpha^b-b^b \label{d2} .
\end{eqnarray}
Let $K^{d+1}=J^{0 \,d+\!1}$. The Lie algebra of $IG(e_{d+1})$ is
spanned by $J_{ab}$,  $W_a$, $P^b$, $H$, $\mathcal{M}$ and
$K^{d+1}$ and has non-vanishing commutations relations given by
Eqs. (\ref{l1}), (\ref{l2}), (\ref{l3}), (\ref{l4}), (\ref{l5})
and
\begin{eqnarray}
{}[K^{d+1},H]&=&H-\mathcal{M} , \\
{}[K^{d+1},\mathcal{M}]&=& -\mathcal{M} , \\
{}[K^{d+1},W_a]&=&-W_a .
\end{eqnarray}
Fortunately $\mathcal{I}=\{ \mathcal{M}\}$ is still an ideal for the
enlarged Lie algebra, and $N=\exp \mathcal{I}$ is a normal subgroup.
The quotient group $IG(e_{d+1})/N$ has a well defined action on the
space $Q$ of light rays $Nx$. The Lie algebra of $IG(e_{d+1})/N$ is
spanned by $\tilde{J}_{ab}=J_{ab}+\mathcal{I}$,
$\tilde{W}_a=W_a+\mathcal{I}$, $\tilde{P}^b=P^b+\mathcal{I}$,
$\tilde{H}=H+\mathcal{I}$, and $\tilde{K}^{d+1}=K^{d+1}+\mathcal{I}$
and satisfies the non-vanishing commutation relations
\begin{eqnarray}
{}[\tilde{J}_{ab},\tilde{J}_{cd}] &=&\delta_{ad}\tilde{J}_{b c}
+\delta_{b d} \tilde{J}_{a
d}-\delta_{a c} \tilde{J}_{b d} -\delta_{b d} \tilde{J}_{a c} , \label{tl1}\\
{}[\tilde{W}_{a}, \tilde{J}_{b c}] &=& \delta_{a b} \tilde{W}_{c}-\delta_{a c} \tilde{W}_{b} , \label{tl2} \\
{}[\tilde{P}_a,\tilde{J}_{b c}]&=&\delta_{a b} \tilde{P}_{c}-\delta_{a c} \tilde{P}_{b}, \label{tl3} \\
{}[\tilde{W}_a, \tilde{H}]&=& \tilde{P}_a, \label{tl4}\\
{}[\tilde{K}^{d+1},\tilde{H}]&=&\tilde{H} ,\\
{}[\tilde{K}^{d+1},\tilde{W}_a]&=&-\tilde{W}_a .
\end{eqnarray}
This is the Lie algebra of the Galilei group plus time dilations
whose action on $Q$ is given by Eqs. (\ref{d1}) and (\ref{d2}).
The boost $K^{d+1}$ on directions $e_{d+1}$ generates time
dilations in the quotient spacetime.

The study of section \ref{obs} on the non-inertial observers can be
generalized. The result is that condition (i) of that section can be
dropped. In order to stay time by time in $S(e_{d+1})$ an observer
must satisfy the following condition. The projections of the angular
velocity vector   and acceleration on the plane perpendicular to
$e_{d+1}$ must be perpendicular to each other and of the same
magnitude, while their vector product must equal $e_{d+1}$ up to a
non-negative factor. Thus, contrary to what happens for the
$S_{\omega}(e_{d+1})$ subset, here the acceleration is not
constrained, since the angular velocity (orientation) can always be
chosen such that the observer belongs time by time to $S(e_{d+1})$.

As a consequence of Eq. (\ref{d1}), if two light spots are
simultaneous in a frame belonging to $S(e_{d+1})$ then they are
simultaneous in every frame in the same set (universality of
Galilean simultaneity). This property gives a way of splitting the
space of null geodesics $Q$ of direction $n$ into subsets of
`simultaneous' null geodesics $S_{t}$ where $t$ is the time of
arrival of the light beam on the screen of a representative observer
in $S(e_{d+1})$. These subsets are the planes $n \cdot x=cnst.$ The
label $t$ is required in order to distinguish between the different
sets but its actual value is not important.

\begin{figure}
\centering  \psfrag{A1}{$x^1$} \psfrag{A2}{$x^2$} \psfrag{A3}{$x^3$}
\psfrag{B1}{${x'}^1$} \psfrag{B2}{${x'}^2$} \psfrag{B3}{${x'}^3$}
\psfrag{K1}{${O}$} \psfrag{K2}{${O'}$} \psfrag{V}{${\bf v}$}
\psfrag{D}{$d$}
\includegraphics[width=7cm]{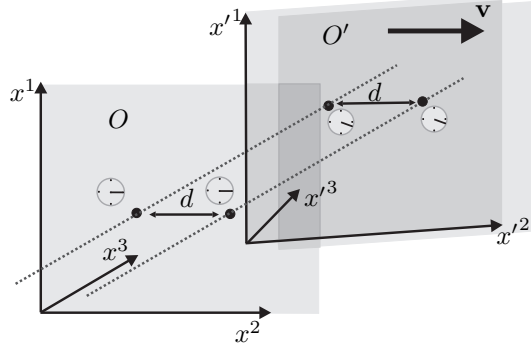}
\caption{The observer $O'$ has velocity ${\bf v}=v {\bf e}_2$ with
respect to $O$ and  is slightly rotated in such a way that for both
frames the light from $\Sigma$ has direction along the third axis.
Consider two light beams from $\Sigma$. If they hit the first screen
at the same time according to $O$'s time then they hit the second
screen at the same time  according to $O'$'s. The Minkowskian
distance $d$ between the two pairs of events is the same which, due
to the previous result, is another way to say that the distance
between the spots on the `screen' $x^3=0$ as measured by $O$ is the
same of the  distance between the spots on the screen ${x'}^3=0$ as
measured by $O'$. The moving screens in this figure are obtained
from the system ${x'}^3=0$ and $x^0=C$, by varying $C$.} \label{fig}
\end{figure}

The Eq. (\ref{d2}) implies that the distance between two
simultaneous light spots does not change under frame changes, that
is, the scalar $x^{a}x_a$ is an invariant. Thus, this scalar gives a
well defined  Euclidean metric on the d-dimensional space of
simultaneous null geodesics of a given direction $S_t$. Figure
\ref{fig} summarizes these results.

\subsection{The shape invariance of shadows} \label{sett} Shadows are
produced by the absence of a light. Where there is a dark spot we
can think that a light beam has been screened by the object that
produces the shadow. Shadows  propagate as light does and thus we
have found a transformation rule for shadows on
perpendicular-to-light screens under inertial frame changes. The
shadow of a pointlike particle worldline is a pointlike shadow
worldline which we can think as composed of a continuous succession
of dark spots.

Some observations are in order:
\begin{description}
 \item(i) Shadows  do not deform their
shape under the change of frame in $S(e_{d+1})$.  This result holds
for non-inertial observers as well as long as they are oriented in
such a way that the light from $\Sigma$ propagates in the direction
of their axis  $e_{d+1}$ (i.e. $K(\tau)$ belongs to $S(e_{d+1})$).
Indeed at any non-inertial observer proper time $\tau$, the
statement holds for the image of the shadow on the screen of
$K(\tau)$ which coincides with that of the non-inertial observer.
Thus, the shape of a shadow can not be deformed by the arbitrary
motion of a orthogonal-to-light screen.
\item (ii) If a shadow changes in time in
an orthogonal-to-light screen of an inertial observer, due for
instance to the motion, rotation or deformation of the object that
projects the shadow, then, up to a Galilean transformation, exactly
the same projected movement is seen by a different inertial observer
with possibly a different time rate if the respective frames are not
related by an inhomogeneous little group transformation.
\item(iii) Since the Galilean group does not
impose an upper bound on the velocities we recover the known fact
\cite[Chap. 6]{schiller03} that shadows can move faster than light
or, in more suggestive terms that ``darkness is faster than
light".
\item (iv) A dark (or light) spot can be at rest in the
screen of a suitable inertial frame as it happens if the object that
projects the shadow is at rest in that same frame (or as it happens
if a screen is opaque and has a hole). Reasoning in the rest frame
one can easily conclude that the shadow of an object in uniform
motion has the shape of the shadow of the object at rest taken with
respect to a suitable direction. In particular the shadow of a
sphere is always a disk (as it happens in the spherical picture case
\cite{penrose59}).
\item (v) As it is well known the attempt at defining what is a ``rigid
motion"  in special relativity leads to several problems. Rigid
motions are particularly restrictive (for instance, Born's rigidity
does not allow to put into rotation a disk which was initially at
rest) and ultimately, all the difficulties arise from the finite
speed of light which makes any kind of ``rigidity" constraint quite
unnatural. Due to the same reason the ``rigid body" does not exist
in special relativity. Nevertheless, given a system of particles and
a preferential lightlike direction $n$, it makes perfect sense to
define its motion as {\em shadow rigid} if the shadows of the
particles preserve their Euclidean distance in time. Due to the
Galilean nature of the transformations, this definition is clearly
independent of the orthogonal-to-light screen chosen.

\end{description}

We recall that the little group of the massless particle
$L(e_{d+1})$ has been investigated in connection with the
classification of elementary particles \cite{weinberg95}. There the
generators $W_a$ were essentially removed from the little group
algebra on the ground that the two degrees of freedom to which they
give rise are not observed.  Weinberg \cite{weinberg64} concludes
that the $ISO(d)$ structure of the little group has no clear
physical significance in particle physics and that the generators
$W_a$ must therefore vanish in the unitary representation. In our
physical problem the generators $W_a$ acquire the clear role of
Galilean boosts for the transformation of shadows between inertial
frames. The role of $W_a$ as `translations' of the little group
$ISO(d)$ expresses the commutativity of Galilean boosts.

\section{The reconstruction problem} \label{ott}
In this section we consider a non-inertial observer whose comoving
inertial frame $K(\tau)$ belongs to $S(e_{d+1})$. Our aim is to
recover the motion of the observer starting from the (observable)
acceleration $\tilde{a}^{d+1}(\tau)$ along the longitudinal
direction $e_{d+1}$, the acceleration $\tilde{a}^a(\tau)$ along the
transverse directions $e_a$ and the angular velocity
$\tilde{\omega}_{a b}(\tau)$ in the plane perpendicular to
$e_{d+1}$. Remarkably, the analogous problem in which the proper
acceleration is decomposed with respect to a
 Fermi-Walker transported tetrad has not yet been solved.
Nevertheless, as we shall see, it can be solved in this case where,
however, the orthonormal basis $\{u,e_{a}, e_{d+1}\}$ is not
Fermi-Walker transported because of the correction needed to
reorient the axis $e_{d+1}$ after an infinitesimal boost. In other
words, whereas the Fermi-Walker transport is a consequence of the
operator $e^{a_i K^{i} \dd \tau+H\dd \tau}$, in our case the
infinitesimal boost (plus time translation) has the form
$e^{(a_{a}W^a +a_{d+1} K^{d+1})\dd \tau+H\dd \tau}$ and therefore
includes an infinitesimal rotation. The transport induced by the
operator $e^{(a_{a}W^a +a_{d+1} K^{d+1})\dd \tau+H\dd \tau}$ will be
called {\em lightlike parallel transport} (relative to the null
direction $n^{\mu}$). A differential geometric characterization of
this parallel transport will be given in appendix \ref{lpt}.

Let $x^{\mu}_{K(\tau)}=G^{\mu}_{\
\nu}[x^{\nu}_{K(0)}-x^{\mu}(\tau)]$ be the transformation from
$K(0)$ to $K(\tau)$. Taking into account the decomposition
$G^{\mu}_{\ \nu}=B^{\mu}_{\ \alpha} S^{\alpha}_{\ \beta}
R^{\beta}_{\ \nu}$, the operator that sends $x^{\mu}_{K(0)}$ to
$x^{\mu}_{K(\tau)}$ can be written
\[
e^{r(\tau) K^{d+1}} e^{\alpha_{a}(\tau)W^a}
e^{\frac{1}{2}\Omega_{ab}(\tau)J^{ab}} e^{t(\tau) H-x^a(\tau) P_a
+x^{d+1}(\tau) \mathcal{M}} ,
\]
where we used the decomposition of $x^{\mu}$, (\ref{dec}). We have
to find an expression for $r(\tau)$, $\alpha_{a}(\tau)$,
$\Omega_{ab}(\tau)$ (or alternatively ${\rm R}^{a}_{\ b}(\tau)$, see
(\ref{rot})), $t(\tau)$, $x^a(\tau) $ and $x^{d+1}(\tau) $ given
$\tilde{a}^{d+1}(\tau)$, $\tilde{a}^a(\tau)$ and
$\tilde{\omega}_{ab}(\tau)$. The infinitesimal transformation of the
observer can be modelled as a (proper) time translation followed by
a transformation in $L(e_{d+1})$
\[
e^{\tilde{a}^{d+1}(\tau) K^{d+1} \dd \tau} e^{ \tilde{a}^a(\tau) W_a
\dd \tau} e^{\frac{1}{2} \tilde\omega_{ab}(\tau) J^{ab} \dd
\tau}e^{H \dd \tau} ,
\]

Note that the space translations do not enter at the infinitesimal
level, as they would imply a violation of causality.

The dependence of $r(\tau)$, $\alpha_{a}(\tau)$,
$\Omega_{ab}(\tau)$, $t(\tau)$, $x^a(\tau) $ and $x^{d+1}(\tau) $ on time is recovered from the constraint
\begin{eqnarray*}
&&[ e^{\tilde{a}^{d+1}(\tau) K^{d+1} \dd \tau} e^{
\tilde{a}^a(\tau) W_a  \dd \tau}  e^{\frac{1}{2}
\tilde\omega_{ab}(\tau) J^{ab} \dd \tau} e^{H \dd \tau} ] [
e^{r(\tau) K^{d+1}} \\ && e^{\alpha_{a}(\tau)W^a}
e^{\frac{1}{2}\Omega_{ab}(\tau)J^{ab}} e^{t(\tau) H-x^a(\tau) P_a
+x^{d+1}(\tau) \mathcal{M}}]\\
&&=e^{r(\tau+\dd \tau) K^{d+1}} e^{\alpha_{a}(\tau +\dd \tau)W^a}
e^{\frac{1}{2}\Omega_{ab}(\tau+\dd \tau)J^{ab}} \\ && e^{t(\tau+\dd
\tau) H-x^a(\tau+\dd \tau) P_a +x^{d+1}(\tau+\dd \tau) \mathcal{M}}
,
\end{eqnarray*}
In order to find  $r(\tau+\dd \tau)$, $\alpha_{a}(\tau+\dd \tau)$,
$\Omega_{ab}(\tau+\dd \tau)$, $t(\tau+\dd \tau)$, $x^a(\tau+\dd
\tau) $ and $x^{d+1}(\tau+\dd \tau)$, we have  to use the
commutation rules of $\mathcal{IG}(e_{d+1})$ several times. The
following observation will be particularly useful. The
Campbell-Baker-Hausdorff formula
\begin{equation}
e^{A}\,e^{B}=e^{A+B+\frac{1}{2}[A,B]+\frac{1}{12}[A,[A,B]]} ,
\end{equation}
 is correct at any order if $[B,[B,A]]=0$ and $[A,[A,B]]$ commutes with $A, B$ and $[A,B]$. This fact implies the identity
\begin{equation}
e^B \,e^A=e^A\, e^B\, e^{-[A,B]} \,e^{\frac{1}{2}[A,[A,B]]} ,
\end{equation}
which is very useful in order to move $e^{H \dd \tau}$ on the right
hand side of  $e^{\alpha_{a}(\tau)W^a}$. Also from the expression
(\ref{gma}) or (\ref{boo}) it can be easily checked that
\[
e^{H \dd \tau} e^{r(\tau) K^{d+1}} = e^{r(\tau) K^{d+1}}
e^{e^{-r(\tau)} H  \dd \tau+\sinh r (\tau) \mathcal{M} \dd \tau} ,
\]
and
\[
e^{\tilde{a}^a(\tau) W_a \dd \tau}  e^{r(\tau) K^{d+1}}= e^{r(\tau)
K^{d+1}}   e^{e^{r(\tau)} \tilde{a}^a(\tau) W_a \dd \tau} .
\]
In the end we arrive at the following set of differential equations
\begin{eqnarray}
\frac{\dd {\rm R}^{a}_{\ b}}{\dd \tau}&=& \tilde{\omega}^{a}_{\ c} {\rm R}^{c}_{\ b}  , \label{s1} \\
\frac{\dd r}{\dd \tau}&=& \tilde{a}^{d+1}, \label{s2}\\
\frac{\dd t}{\dd \tau}&=& e^{-r} , \label{s3} \\
\frac{\dd \alpha^a}{\dd \tau}&=& e^r \tilde{a}^a+\tilde{\omega}^{a}_{ \ b}\alpha^b   ,  \label{s4}\\
\frac{\dd x^a}{\dd \tau}&=&e^{-r} \alpha_{b} {\rm R}^{b a}   , \label{s5}\\
\frac{\dd x^{d+1}}{\dd \tau}&=&   \sinh r+ \frac{e^{-r}}{2}
\alpha^a \alpha_a . \label{s6}
\end{eqnarray}
The lightlike parallel transport is obtained for
$\tilde{\omega}^{a}_{\ c}=0$. Note that each one of the first three
equations is independent of the others. It can be easily checked
that
\begin{equation} \label{uuu}
u^{\mu}=\frac{\dd x^{\mu}}{\dd \tau}=\!\!\begin{pmatrix}
\cosh r+ \frac{e^{-r}}{2}
\alpha^a \alpha_a  \\
 e^{-r} \alpha_{b} {\rm R}^{b a} \\
 \sinh r+ \frac{e^{-r}}{2}
\alpha^a \alpha_a
\end{pmatrix} ,
\end{equation}
and $u^{\mu} u_{\mu}=-1$. Thus $x^{\mu}(\tau)$ is timelike as
expected.

An interesting  consequence of Eq. (\ref{s3}) is that if, for any
$\tau$, $K(\tau) \in S_{\omega}(e_{d+1})$ for a suitable $\omega$,
then $r=0$ and $t=\tau$, that is the proper time of  $K(\tau)$'s
origin coincides with the proper time of its shadow on $K(0)$'s
screen.

\subsection{Absolute transverse orientation} \label{ott2}
The equation (\ref{s1}), although with a different interpretation,
has been studied in many references \cite{levi96}. Fortunately, it
can be completely integrated in the 4-dimensional spacetime case,
since $d=2$. Let $\tilde\omega^{\mu}$ be  the angular velocity
vector
\begin{displaymath}
{\rm R}^{a}_{\ b}(\tau)=\!\!\begin{pmatrix}
\cos \theta(\tau) & \sin \theta (\tau) \\
 -\sin\theta(\tau)& \cos \theta(\tau)
\end{pmatrix}\!\!, \ \ \omega^{a}_{\ b}(\tau)=\tilde\omega^{d+1}\!(\tau)\!\begin{pmatrix}
\!0 & \!1  \\
 \!-1& \!0
\end{pmatrix}\!\!,
\end{displaymath}
then $\theta(\tau)=\int_{0}^{\tau} \tilde\omega^{d+1}(\tau)\dd
\tau$. More generally the fact that Eq. (\ref{s1}) is independent of
the other equations implies that ${\rm R}^{a}_{\ b}(\tau)$ (and
hence $R^{\mu}_{\ \nu}(\tau)$) is independent of the acceleration
history of the frame. This feature is due to the existence of an
absolute transverse orientation induced by the lightlike parallel
transport. In other words assume that $\tilde{\omega}^{a}_{\ c}=0$,
then if the frame $K(0)$ can be moved staying time by time into
$S(e_{d+1})$ to $K(\tau)$ it can not be moved to any other frame
$K'(\tau)$ which differs from $K(\tau)$ only for a rotation which
keeps $e_{d+1}$ fixed. The unique orientation of $K(\tau)$,
dependent on the choice of initial orientation of $K(0)$, and the
arbitrariness in the origin of coordinates of $K(\tau)$, implies an
absolute transverse orientation on spacetime.

To see this consider the following argument. On the set $S(e_{d+1})$
(resp. $S_{\omega}(e_{d+1})$) acts freely the subgroup
$IGP(e_{d+1})$ generated by $K^{d+1}$, $W_a$, $H$, $\mathcal{M}$,
$P_a$ (resp. $ILP(e_{d+1})$ generated by the same set without
$K^{d+1}$). The quotient space is isomorphic to the group $SO(d)$.
Thus the set $S(e_{d+1})$ (resp. $S_{\omega}(e_{d+1})$) splits into
classes $S_{{\rm R}} (e_{d+1})$ (resp. $S_{{\rm R},\omega}
(e_{d+1})$) where the abstract symbol $R$ belongs to the quotient
space isomorphic to $SO(d)$ and denotes the absolute orientation of
the frame induced by the lightlike parallel transport. An observer
belonging to $S_{{\rm R}} (e_{d+1})$ can not move, at a later proper
time, to a frame belonging $S_{{\rm R'}} (e_{d+1})$, $R'\ne R$, if
over it does not act an operator which does not belong to
$IGP(e_{d+1})$ (resp. $ILP(e_{d+1})$), i.e. if in the mean time
$\tilde\omega_{ab}=0$.

The following fact can be easily checked using
Eq.(\ref{s1})-(\ref{s6}). Let
\[\{ {\rm R'}^{a}_{\ b}, r', t',
{\alpha'}^a,{x'}^a,{x'}^{d+1}\}(\tau)\]
 be a solution of the above
equations given $\{ \tilde{\omega}'_{a b},\tilde{a}{'}^{
a},\tilde{a}{'}^{d+1}\}(\tau)$, then
\[\{ {\rm R}^{a}_{\
b}=\delta^{a}_{b}, r=r', t=t', {\alpha}^a={\alpha'}_b{\rm R'}^{b
a},{x'}^a={x}^a,{x}^{d+1}={x'}^{d+1}\}(\tau)\]
 is a solution of the
above equations given $\{ \tilde{\omega}_{a b}=0,\tilde{a}{}^{
a}=\tilde{a}{'}_{b}{\rm R'}^{b a}
,\tilde{a}^{d+1}=\tilde{a}{'}^{d+1}\}(\tau)$. It follows that the
integration of the system reduces to equation (\ref{s1}), that can
be completely solved if $d=2$, and  to the system
(\ref{s2})-(\ref{s6}) with $\omega_{ab}=0$, $R^{a}_{\
b}=\delta^{a}_{b}$ which describes a lightlike transported frame.

\subsection{Integration of the lightlike parallel transport and spacetime
navigation} \label{ott3}

 We integrate the system in the case of a
lightlike transported frame, i.e. $\omega_{ab}=0$, $R^{a}_{\
b}=\delta^{a}_{b}$. Since for $\tau=0$ the frame of the
non-inertial observer in $S(e_{d+1})$ coincides with $K(0)$ we
have, to begin with,
\begin{eqnarray}
r(\tau)&=&\int_{0}^{\tau}\tilde{a}^{d+1}(\tau') \dd \tau', \label{lk1}\\
t(\tau)&=&\int_{0}^{\tau} e^{-r(\tau')} \dd \tau' , \\
\alpha^{a}(\tau)&=& \int_{0}^{\tau} e^{r(\tau')} \tilde{a}^a(\tau')
\dd \tau'  \label{lk3}.
\end{eqnarray}
Eqs. (\ref{s3}) and (\ref{s5}) imply
\begin{equation}
\frac{\dd x^{a}}{\dd t}=\alpha^a ,
\end{equation}
that is, the shadow of $K(\tau)$'s origin   has a velocity on
$K(0)$'s screen which  equals the group parameter $\alpha^a$.
Moreover, if $K(\tau) \in S_{\omega}(e_{d+1})$ for any $\tau$ then
$r=0$ and using Eqs. (\ref{s3}) and (\ref{s4})
\begin{equation}
\frac{\dd \alpha^{a}}{\dd t}=\tilde{a}^a ,
\end{equation}
which means that the acceleration of the shadow  on the screen
equals the actual transverse acceleration measured by the
non-inertial observer using comoving accelerometers.

In general any timelike worldline $x^{\mu}(\tau)$ starting at the
origin of $K(0)$ can be  interpreted as the motion of the origin of
a lightlike transported non-inertial frame. Given $x^{\mu}(\tau)$
one can calculate $u^{\mu}(\tau)$, hence $r(\tau)=-\ln
(u^{0}-u^{d+1})$, $\alpha_a(\tau)=e^{r} u_a $ and finally the
longitudinal and transversal acceleration measured by the
non-inertial observer through Eqs. (\ref{s2}) and (\ref{s4}). The
inverse problem of obtaining the trajectory from the measured
acceleration is the reconstruction problem.

The worldline $x^{\mu}(\tau)$ can be obtained  without
difficulties from Eqs. (\ref{uuu}), (\ref{lk1}) and (\ref{lk3}).
The provided analytical solution of the reconstruction problem
could be applied in futurable spacetime navigation. Far from the
massive sources of gravity the spacetime is almost Minkowskian. A
method that gives to the non-inertial observer a way to recover
its own inertial coordinates would be welcome as in this way a
direct communication of the observer  with an inertial observer
$K(0)$ would be avoided. Indeed, as the distance between the
observers increases, a communication between them becomes
unlikely. In any case, such communication would introduce a delay
that would practically forbid an autonomous spacetime navigation,
and hence the possibility of correcting the trajectory in short
decisional times. The formulas given here allow an autonomous
navigation of the non-inertial observer (spaceship).

The comoving laboratory should be provided with an accelerometer and
three  orthogonal gyroscopes that time by time correct their own
orientation so that the last one (the direction $e_{d+1}$) points
always towards the opposite direction of a given star in the night
sky sphere (the light from the star determines the null direction
$n$). The correction must consist in infinitesimal rotations along
axes perpendicular to $e_{d+1}$. If not corrected the gyroscopes
would give a Fermi-Walker transported frame. In other words a
suitable onboard instrumentation can reproduce without difficulties
a lightlike transported frame (coincident up to a reorientation to
the comoving Fermi-Walker transported triad). The measured
acceleration can therefore be projected on the gyroscopic directions
to obtain $\tilde{a}^{d+1}(\tau)$ and $\tilde{a}^{b}(\tau)$. Given
the acceleration history, the non-iniertial observer can recover the
coordinates $x^{\mu}(\tau)$ by integration using the above formulas.
As far as we know, the analytical solution to the reconstruction
problem provided here represents the only one available for the case
$d
>0$.

\subsection{Motion reconstruction  from  projection and meaning of the
action} \label{ott4}

In this subsection we consider another  kind of reconstruction
problem. This time we assume to have given the shadow motion of
$K(\tau)$'s origin on $K(0)$'s screen and $r(t)$ (or alternatively
the frequency $\frac{\omega'(t)}{\omega(0)}=e^{-r(t)}$). Then,
$K(\tau(t))$'s origin in $K(0)$'s coordinates is (with a dot we
denote the differentiation with respect to $t$)
\begin{equation} \label{xxx}
x^{\mu}(t)=\!\!\begin{pmatrix}
\int_{0}^{t} \frac{e^{2r(t')}+1}{2} \dd t' +\int_{0}^{t}\frac{1}{2} \dot{x}^a \dot{x}_a \dd t'\\
 x^{a}(t)\\
 \int_{0}^{t} \frac{e^{2r(t')}-1}{2} \dd t'+\int_{0}^{t}\frac{1}{2} \dot{x}^a \dot{x}_a \dd t'
\end{pmatrix}.
\end{equation}
This equation gives a meaning to the classical action as it shows
that $\int_{0}^{t}\frac{1}{2} \frac{\dd x^a}{\dd t}\frac{\dd
x^a}{\dd t} \dd t'$ represents the coordinate $x^{d+1}(t)$ of
$K(\tau)$'s origin if $K(\tau) \in S_{\omega}(e_{d+1})$ (hence
$r=0$), given the shadow worldline $x^a(t)$ of $K(\tau)$'s origin.

\subsection{Differential aging from acceleration} \label{ott5}

In this section we generalize to arbitrary  spacetime dimensions
$(d+1)+1$, a formula for the differential aging in terms of the
acceleration history given by the author in \cite{minguzzi04c} for
the $d=0$ case. By differential aging $\Delta$ we mean the
difference between the proper time $T$ needed by an inertial
observer to go from $x^{\mu}(0)$ to $x^\mu(\tau)$ and the proper
time $\tau$ of the non-inertial observer that follows the trajectory
$x^{\mu}(\tau)$. It is well known that $T(\tau) \ge\tau$ where the
equality holds iff $x^{\mu}(\tau)$ is a straight line, however, an
explicit formula for $T$ or $\Delta=T-\tau$, in terms of the
acceleration history is in general difficult to obtain. The above
results lead to
\begin{eqnarray}
T^2-\tau^2&=& \!\!\! \sum_{a=1}^{d} \{ [\int_{0}^{\tau}
e^{-r(\tau')}\dd \tau'] [\int_{0}^{\tau} e^{-r(\tau')}\alpha^a
\alpha^a \dd \tau'] -[\int_{0}^{\tau} e^{-r(\tau')} \alpha^a \dd \tau']^2 \} \nonumber\\
&&\!\!\!+ \{[\int_{0}^{\tau} e^{-r(\tau')}\dd \tau'][\int_{0}^{\tau}
e^{r(\tau')}\dd \tau']-\tau^2 \} ,
\end{eqnarray}
where $r$ and $\alpha$ are given by Eqs. (\ref{lk1}) and
(\ref{lk3}). We did not use the sum-over-repeated-indexes convention
to point out that the $i$-th term between braces on the right-hand
side is, thank to the Cauchy-Schwarz inequality, greater than zero
and vanishing iff $\tilde{a}^i(\tau')=0$ for every $\tau' \in
[0,\tau]$. This observation leads to the expected inequality $T\ge
\tau$, where the equality holds iff the acceleration vanishes. If
$d=0$ the right hand side reduces to the last term and the formula
reduces the the differential aging formula given in
\cite{minguzzi04c}.

The expression for $T$ given the shadow worldline $x^{a}(t)$ and
$r(t)$ reads instead
\begin{equation}
 T^2=t\int_0^t \dot{x}^a \dot{x}_a \dd t'-x^a
x_a + t \int_{0}^{t} e^{2r(t')}\dd t' .
\end{equation}

\subsection{Position and frequency drifts} \label{ott6}
The last component of the non-inertial observer  worldline $x^{\mu}(\tau)$ is
\begin{eqnarray}
\!\!\!\!\!\!x^{d+1}(t)\!\!&=&\!x^{d+1}(0)+\frac{1}{2}\int_{0}^{t}[e^{2r(t')}-1+ \dot{x}^{a}\dot{x}_{a}] \dd t' \nonumber\\
\!\!\!\!\!\!\!\!\!&=&
\!x^{d+1}(0)+\frac{1}{2}\int_{0}^{t}[(\frac{\omega}{\omega'(t)})^2-1+
\dot{x}^{a}\dot{x}_{a}] \dd t', \label{xd}
\end{eqnarray}
where $\omega$ is the frequency of light going in direction
$e_{d+1}$ as measured in the  inertial frame $K$ and $\omega'(t)$ is
the frequency measured by the non-inertial observer. The relevance
of this equation comes from its unexpected consequences. For
instance, assume that the non-inertial observer stays very close to
the last axis of $K$, so that $0 \simeq \vert x^a \vert \ll \vert
x^{d+1}\vert$, and the distance is given by $\vert x^{d+1}\vert$
with a small error. Moreover, assume that $\omega'=\omega$. If
$x^{d+1} <0$  the signal of frequency $\omega$ could be thought as
emitted by the non-inertial observer. The inertial observer measures
the same frequency and  in practice knows quite accurately the
position of the non-inertial observer up to the value of coordinate
$x^{d+1}$. From the fact that the frequency does not change her
could be tempted to infer that $x^{d+1}$ is constant in time as a
velocity component along the axis would imply a Doppler effect.
However, the previous formula gives
\[
x^{d+1}(t)=x^{d+1}(0)+\frac{1}{2}\int_{0}^{t}\dot{x}^{a}\dot{x}_{a}
\dd t' ,
\]
which means that, due to the transversal motion, the position may
drift along the last  axis without any frequency change. In
particular the drift is such that the coordinate $x^{d+1}$ can
only increase. Thus the knowledge of the transversal position of
the non-inertial observer does not place a bound on the
transversal velocities which are at the origin of the effect.
Remarkably, transversal velocities are difficult to measure and
hence the effect may be present in practical applications.

A related effect takes place if the second non-inertial observer is
positioned near  the last axis of $K$, i.e.  $0 \simeq \vert x^a
\vert \ll \vert x^{d+1}\vert$ at a fixed value of the coordinate
$x^{d+1}$. Assume that $x^{d+1}>0$, so that $\omega$ can be
identified with the frequency of a wave emitted by $K$ and received
by the non-inertial observer. It could be naively expected that the
measured frequency $\omega'$ is  constant since the non-inertial
observer is almost at rest, but differentiation of Eq. (\ref{xd})
gives
\[
\omega'(t)=\frac{\omega}{\sqrt{1-\dot{x}^{a}\dot{x}_{a}(t)}} .
\]
The transversal velocity can change the frequency without affecting the average position.
%

\section{From Poincar\'e to Galilei invariance: particle
collisions}\label{fp}

If a Poincar\'e invariant physical phenomenon in $(d+1)+1$-Minkwoski
spacetime is projectable on $Q$ then the projected phenomenon is
Galilei invariant \cite{elizade78}. Indeed, the original physical
phenomenon is invariant under the Poincar\'e subgroup $IL({\bf
e}_{d+1})$ which means that the projection is invariant under the
Galilei group $Gal(d)$. The aim of these last sections is to show
explicitly the Galilei invariance of the shadow of a relativistic
collision.

We choose an arbitrary inertial observer in $S_{\omega}({e}_{d+1})$
of coordinates $\{x^{\mu}\}$ and consider a particle of worldline
$x^{\mu}(\lambda)$ and momentum $p^{\mu}=\dd x^{\mu}/\dd \lambda$.
 We are interested on the
lightlike projection $x^a(t)$ of the curve $x^{\mu}(\lambda)$ on the
screen $x^{d+1}=0$, where the projected curve is parametrized with
the Galilean time $t$.  The velocity of the shadow on the screen is
\begin{equation} \label{ve}
\dot{x}^a=\frac{\dd x^0}{\dd t}
v^a=\frac{v^a}{1-v^{d+1}}=\frac{p^a}{p^0-p^{d+1}} .
\end{equation}

Let us temporarily focus on a massive particle on $M$ and then
generalize the results to the massless case.  The equation of motion
for the free particle of mass $m$ is obtained from Hamilton's
principle in configuration space (see Eq. (\ref{mink}))
\begin{equation}
0=\delta \!\!\int\!\! m \dd
\tau=\delta\!\!\int_{t_0}^{t_1}\!\!\!\!\! m
\sqrt{2\dot{x}^{d+1}+1- \dot{\bf x}^2} \,\dd t .
\end{equation}
The momentum conjugated to $x^{d+1}$ is the {\em shadow mass}
\begin{eqnarray}
\tilde{m}&=&-n^{\mu}p_{\mu} =p^0-p^{d+1} . \label{h}
\end{eqnarray}
The last expression  makes sense also for massless particles on
$M$ and it is invariant under inertial frame changes generated by
$IL({\bf e}_{d+1})$. The cyclic variable $x^{d+1}$ can be removed
by using Routh's reduction \cite[Sect. 8.9]{marsden99}. On the
reduced spacetime $Q$ Hamilton's action principle holds where the
new Lagrangian is replaced by the Routhian, i.e. a suitable
Legendre transform of the original Lagrangian  in which the
conjugated momentum $\tilde{m}$ is regarded as a constant. In our
case the Routhian is
\[
\frac{\tilde{m}}{2} \dot{\bf x}^2-
\frac{1}{2}[\frac{m^2}{\tilde{m}}+\tilde{m} ].
\]
From this expression we deduce the shadow kinetic energy $T$,
which has the same form as in classical mechanics, and the shadow
 internal energy $I$
\begin{eqnarray}
T&=&\frac{\tilde{m}}{2} \dot{\bf x}^2 ,\\
I&=&\frac{1}{2} [\frac{m^2}{\tilde{m}}+\tilde{m}] . \label{pot}
\end{eqnarray}
With minor differences these identifications can also be found in
\cite{bouchiat71}. The internal energy depends on a parameter $m$
which has no (shadow) kinematical interpretation. Indeed, in
classical non-relativistic mechanics the internal energy of a
particle can not be expressed in terms of the mass alone as in
special relativity. For instance, a body $A$ composed by  bodies
$B$, $C$ and a compressed spring has a total mass given by
$m_A=m_B+m_C$. Nevertheless, if the particle $A$ separates into
$B$ and $C$,  the potential energy
 of the spring must be taken into account in the energy balance.
It is therefore natural that a non-kinematical parameter $m$
enters in the expression of the shadow internal energy. The fact
that the above identification of kinetic and internal energy is
correct will be confirmed in a moment when we shall study the
conservation of energy.

Note that in any case the potential energy is a constant and
therefore it is irrelevant in the variational principle. We conclude
that the reduced action coincides with the classical action with the
{\em shadow masses} provided by Eq. (\ref{h}).

The shadow mass $\tilde{m}=p^0-p^{d+1}$ is positive unless the
particle on $M$ is massless with momentum of direction $n^{\mu}$, in
which case it vanishes. This fact is coherent with the previous
observation that those massless particles on $M$ do not represent
particles on $Q$ but, rather, events. Analogously, the definitions
of shadow kinetic energy and shadow internal energy generalize to
massless particles on $M$. Indeed they can be directly expressed in
terms of the covariant momentum $p^{\mu}$
\begin{eqnarray}
T&=&\frac{\tilde{m}}{2} \dot{\bf x}^2= \frac{p^{a}p_{a}}{2(p^0-p^{d+1})},\\
I&=&\frac{1}{2} [\frac{m^2}{\tilde{m}}+\tilde{m}] =\frac{1}{2}
[\frac{m^2}{p^{0}-p^{d+1}}+p^0-p^{d+1} ] .
\end{eqnarray}
The shadow momentum  can be obtained from Eqs. (\ref{h}) and
(\ref{ve})
\begin{equation} \label{grt}
\tilde{m}\dot{x}^a=p^a .
\end{equation}
The energy reads
\begin{eqnarray}
E\!\!&=&\!T+I\!=\!\!\frac{1}{2}\frac{p^a
p_{a}}{p^0-p^{d+1}}+\frac{1}{2}[\frac{m^2}{p^0-p^{d+1}}+p^0-p^{d+1})]=p^0
. \label{hgf}
\end{eqnarray}

\subsection{Conservation of shadow mass, momentum and energy}
\label{und}

Let us now consider a collision of $N$ particles  of momentum
$p^{\mu}_{(i)}$, $i=1,\ldots, N$. From the collision $N'$ final
particles of momentum ${p'}^{\mu}_{(j)}$, $j=1,\ldots, N'$,
emerge. The conservation of momentum in Minkowski spacetime reads
\begin{equation} \label{con}
\sum_{i=1}^N p^{\,\mu}_{(i)}= \sum_{i=1}^{{N'}} {p'}^{\,\mu}_{(i)}
.
\end{equation}
The particle worldlines are geodesics that project into geodesics
of $Q$: the shadow worldlines. These straight lines in $Q$
collide. Subtracting Eq. (\ref{con}) for $\mu=0$ and $\mu=d+1$ we
obtain
\begin{equation} \label{mass}
\sum_{i=1}^N \tilde{m}_{(i)}= \sum_{i=1}^{{N'}} {\tilde{m}'}_{(i)}
,
\end{equation}
that is, the total shadow mass is conserved in the shadow
collision. From Eq. (\ref{con}) with $\mu=1,\ldots,d$ we obtain
\begin{equation} \label{mome}
\sum_{i=1}^N \tilde{m}_{(i)}\dot{{\bf x}}_{(i)}= \sum_{i=1}^{{N'}}
{\tilde{m}'}_{(i)}{\dot{{\bf x}}'}_{(i)} ,
\end{equation}
that is, the shadow momentum is conserved. The found conservation
principles are characteristic of a non-relativistic system. Indeed
they are consequence of the underlying Galilean symmetry.

Let us come to the conservation of energy. Since the relativistic
energy is conserved, from Eq. (\ref{hgf}) we deduce that the
shadow energy is conserved
\begin{equation} \label{ener}
\sum_{i=1}^{N} E_{(i)}=\sum_{i=1}^{{N'}} {E'}_{(i)} .
\end{equation}
In order to prove the conservation of shadow mass, momentum and
energy we  used four independent momentum conservation equations
on $M$. Thus the obtained shadow conservation principles are
equivalent to the original ones given by  Eq. (\ref{con}).
However, there is an important difference. The internal energy in
the classical case is not a kinematical observable contrary to
what happens in the relativistic case where the energy of the
particle can be inferred from the mass and the covariant velocity.
In some interesting cases this unknown internal energy plays no
role in the shadow collision.

\subsection{Conservation of shadow kinetic energy and massless
particles} \label{und2}

 We give some definitions. An {\em elastic}
relativistic collision in $M$ is a collision (i.e. Eq. (\ref{con})
holds) in which the particle species are conserved, $N={N'}$ and
$m_{(i)}={m'}_{(i)}$, $i=1,\ldots,N$. An {\em elastic classical}
collision in $Q$ is a collision (i.e. Eqs. (\ref{mass}) and
(\ref{mome}) hold) in which the kinetic energy is conserved
\begin{equation}
\sum_{i=1}^{N} \frac{\tilde{m}_{(i)}}{2}\dot{{\bf x}}_{(i)}^2 =
\sum_{i=1}^{{N'}} \frac{{\tilde{m}'}_{(i)}}{2}{\dot{{\bf
x}}}{'}^{2}_{\!(i)} .
\end{equation}
We shall say that a classical elastic collision is {\em proper} if
the particle  species are conserved $N={N'}$,
$\tilde{m}_{(i)}={\tilde{m}'}_{(i)}$, $i=1,\ldots,N$ and the total
kinetic energy is conserved

From the expression of the shadow internal energy we obtain that
{\em the shadow of an elastic relativistic collision is a proper
elastic classical collision provided the shadow masses are
preserved in the collision}, $\tilde{m}_{(i)}={\tilde{m}'}_{(i)}$,
$i=1,\ldots,N$.

Finally, let us consider the shadow of a collision in which the
initial and final particles are massless. Since the sum of two
non-collinear null vectors is a timelike vector, a collision of
this form must involve at least two initial and two final
particles. From the expression of the internal energy,
$I=\tilde{m}/2$, we conclude that {\em  the shadow of a collision
that involves only massless particles is a classical elastic
collision}.

\subsection{The inverse problem} \label{dod}

Suppose we are given the data of a collision on $d+1$ Galilean
spacetime. In this section we want to construct a collision in
(d+1)+1 Minkowski spacetime such that the original collision can
be regarded as its shadow. Thus  assume that we are given numbers
$N$, ${N'}$, of initial and final particles, their (positive)
masses $\tilde{m}_{(i)}$ and ${\tilde{m}'}_{(i)}$ and their
velocities $\dot{\bf x}_{(i)}$, $\dot{{{\bf x}}}'_{(i)}$. Assume
the conservation of total mass
\begin{equation}
\sum_{i=1}^N \tilde{m}_{(i)} = \sum^{{N'}}_{i=1}
{\tilde{m}'}_{(i)} ,
\end{equation}
and momentum
\begin{equation}
\sum_{i=1}^N \tilde{m}_{(i)} \dot{\bf x}_{(i)} = \sum^{{N'}}_{i=1}
{\tilde{m}'}_{(i)} \dot{{{\bf x}}}'_{(i)} .
\end{equation}
We want to  lift the collision to $M$ in a way
 invariant under Galilei transformations, i.e. the lift to $M$
must depend only on the collision on $Q$ and not on the choice of
Galilean frame on $Q$. We known that the kinetic energy $T_{(i)}$
of particle $(i)$ and the  total kinetic energy $T=\sum^{N}_{i}
T_{(i)}$, are not invariant under Galilean transformation.
Nevertheless, the difference between final and initial kinetic
energies is invariant
\begin{equation} \label{csq}
\Delta T={T'}-T=\sum^{{N'}}_{i=1} \frac{1}{2}{\tilde{m}'}_{(i)}
\dot{{{\bf x}}}{'}^2_{\!\!(i)} -\sum^{{N}}_{i=1}
\frac{1}{2}\tilde{m}_{(i)} \dot{\bf x}_{(i)}^2 .
\end{equation}
This fact can be easily checked using the conservation of Galilean
mass and momentum assumed above. One should be careful because
$\Delta T_{(i)}$ is not Galilei invariant. The quantity $\Delta T$
and the masses $\tilde{m}_{(i)}$, ${\tilde{m}'}_{(i)}$, do not
depend on the Galilean frame chosen. Assume that constants
$m_{(i)}$ and ${m'}_{(i)}$ are given such that
\begin{equation} \label{cde}
2\Delta T= \sum^{{N}}_{i} \frac{{m}_{(i)}^2}{\tilde{m}_{(i)}}
-\sum^{{N'}}_{i} \frac{{m'}_{\!(i)}^2}{{\tilde{m}}'_{(i)}}.
\end{equation}
By giving these numbers one fixes the (Galilei invariant) internal
energy of the particles involved in the collision on $Q$ according
to the formula (\ref{pot}). Clearly with this definition of
$I_{(i)}$ the total energy is conserved
\begin{equation}
\sum_{(i)}^{N} E_{(i)}=\sum_{(i)}^{\bar{N}} \bar{E}_{(i)} .
\end{equation}
Then the collision on $Q$ is the shadow of a collision in (d+1)+1
Minkowski spacetime constructed as follows. The particles have
masses $m_{(i)}$ (${m'}_{(i)}$) and covariant momentum
\begin{eqnarray*}
p^{\mu}_{(i)}&=&\begin{pmatrix} E \\
\tilde{m} \dot{\bf x} \\
E-\tilde{m}
\end{pmatrix}_{\!\!(i)}\!\!\! =
(E-\tilde{m})_{(i)} n^{\mu} + \begin{pmatrix}
\tilde{m} \\
\tilde{m} \dot{\bf x} \\
0
\end{pmatrix}_{\!\!(i)}\!\!
=\frac{\tilde{m}}{2} \begin{pmatrix} \dot{\bf x}^2+
\frac{m^2}{\tilde{m}^2}+1 \\
2 \dot{\bf x} \\
 \dot{\bf x}^2+ \frac{m^2}{\tilde{m}^2}-1
\end{pmatrix}_{\!\!(i)}\!\!\! ,
\end{eqnarray*}
such that $p^{\mu}_{(i)}p_{\mu (i)}=-m_{(i)}^2$. With these
definitions the total momentum is conserved. Now, we have to give
some more data in order to fix the motion of the particles in $M$.
Since we have fixed the momentum of the particles we have only to
give the event of collision. It must be chosen in the null
worldline that projects on the event of collision in $Q$. This
gives one more real parameter. Once fixed the motion of the
particles is determined both forward and backward in time with
respect to the instant of collision.

Particularly interesting are the elastic collisions on $Q$, i.e.
$\Delta T=0$. In this case, up to translations generated by
$\mathcal{M}$, there is a canonical $\frac{1}{\mu}$-lift from $Q$
to $M$, $\frac{1}{\mu} \in [0,+\infty)$ obtained by setting
$m_{(i)}=\frac{1}{\mu} \tilde{m}_{(i)}$ and analogously for
${m'}_{(i)}$. Indeed with this choice, due to the conservation of
(shadow) mass, the constraint (\ref{cde}) is satisfied. The
internal energy becomes
$I_{(i)}=\frac{1+1/\mu^2}{2}\tilde{m}_{(i)}$, and (restoring $c$)
we find that for $\mu=1$  it takes the usual relativistic form,
$\tilde{m}_{(i)} c^2$.


There are two natural choices for $1/\mu$
\begin{description}
\item[Lightlike 0-lift, $\frac{1}{\mu}=m_{(i)}={m'}_{\!\!(i)}=0$].
{\em The elastic collision on $Q$ can be regarded as the shadow of
a collision on $M$ in which only massless particles  of momentum}
\begin{equation}
p^{\mu}_{(i)}=\frac{\tilde{m}_{(i)}}{2}\begin{pmatrix} \dot{\bf
x}^2+1\\
2 \dot{\bf x} \\
\dot{\bf x}^2-1
\end{pmatrix}_{\!\!(i)}\!\!\! ,
\end{equation}
{\em are involved}.
\item[Timelike 1-lift, $\frac{1}{\mu}=1$, $m_{(i)}=\tilde{m}_{(i)}$,
${m'}_{(i)}={\tilde{m}'}_{(i)}$]. {\em The elastic collision on
$Q$ can be regarded as the shadow of a collision between particles
 of momentum}
\begin{equation}
p^{\mu}_{(i)}=\frac{\tilde{m}'_{(i)}}{2}\begin{pmatrix} \dot{\bf
x}^2+2\\
2 \dot{\bf x} \\
\dot{\bf x}^2
\end{pmatrix}_{\!\!(i)}\!\!\! ,
\end{equation}
{\em for which the shadow mass coincides with the mass}.
\end{description}

\section{Conclusions} \label{conc}

Some classical aspects of lightlike dimensional reduction have
been studied.
The Galilean transformation property of shadows represents surely
the most intuitive way to grasp the underlying mathematics. As an
interesting consequence it gives to the Galilei group in 2+1
dimensions the status of exact physical symmetry once we agree
that it should be applied to the transformation of shadows rather
than to events (at least in absence of curvature otherwise even
the Poincar\'e group is broken).

Also, the emphasis made on the role of observers on the full
spacetime $M$ allowed us to recognize the usefulness of lightlike
dimensional reduction for autonomous spacetime navigation. An
important role was played by the concept of lightlike parallel
transport which we introduced twice, using a group theoretical
definition or an equivalent differential geometric definition (in
the appendix).


In the last section we studied in detail the shadow of a
relativistic collision. Apart from results that could have been
expected in view of  the existence of a Galilei quotient subgroup
inside the Poincar\'e group, a further interesting result was
obtained which relates
 the masslessness of the particles involved in the collision to
the conservation of kinetic energy in the projected shadow
collision.

\section*{Acknowledgments} I thank C. Tejero Prieto for suggesting
the analogy between this work and Plato's allegory of the cave, and
M. Santander and M. Modugno for pointing out some works on,
respectively,  front wave dynamics and Newton-Cartan theory. I also
would like to thank M. S\'anchez for its accurate reading of the
manuscript and for the many suggestions. This work has been
partially supported by INFN, Grant No. 9503/02.

\appendix

\section{Lightlike parallel transport} \label{lpt} In this
appendix we give a differential geometric characterization of the
lightlike parallel transport  already introduced in section
\ref{ott} by group theoretical means. The results of this section
hold as well in a curved spacetime having a covariantly constant
null vector field $n$, $n_{\mu;\nu}=0$ (the direction of light).

Consider a non-inertial observer on $M$, that is a timelike
worldline $\gamma(\tau)$ parametrized with respect to proper time,
and an orthonormal frame $\{u=\p_{\tau},e_{i}\}$ along it. The
Fermi-Walker derivative $\nabla^{FW}_{u}$ is a minimal modification
of the covariant derivative $\nabla_{u}$, such that
$\nabla^{FW}_{u}u=0$. A vector field $v^{\mu}(\tau)$ along $\gamma$
is Fermi-Walker transported, $\nabla^{FW}_{u}v=0 $ iff its
components with respect to the Fermi-Walker transported tetrad $\{u,
e^{FW}_i\}$, $\nabla^{FW}_{u} e^{FW}_i=0$ do not change. As it is
well know the condition of preserving the orthonormality, gives, for
a generic $v$,
\begin{equation}
\nabla_{u}v_{\mu}-\nabla^{FW}_{u}v_{\mu}=\Omega^{FW}_{\mu
\nu}v^{\nu},
\end{equation}
where $\Omega^{FW}_{\mu \nu}$ is a 2-form on $\gamma$. The minimal
modification which leads to $\nabla^{FW}_{u}u=0$ is then the
choice
\begin{equation}
\Omega^{FW}_{\mu \nu}=u_{\mu}a_{\nu}-u_{\nu}a_{\mu}.
\end{equation}
Analogously we are looking for a lightlike covariant derivative
$\nabla^{L}_{u}$ along $\gamma$  which measures the variation of
the components of a vector field $v$ with respect to the tetrad
$\{u,e^L_{i}\}$ of a lightlike transported frame. Here the
lightlike transported frame is such that the null vector field $n$
has always the same components up to a factor that may change in
time, i.e. the null vector field $n$ has always the same direction
with respect to a lightlike transported frame. Then, with respect
to the lightlike transported frame the  vector
$n^{\mu}/(n^{\beta}u_{\beta})$ has constant components and hence
the further condition which defines the lightlike transported
frame is f $\nabla^{L}_{u}[n^{\mu}/(n^{\beta}u_{\beta})]=0$. Let
us write
\begin{equation}
\nabla_{u}v_{\mu}-\nabla^{L}_{u}v_{\mu}=\Omega^{L}_{\mu
\nu}v^{\nu},
\end{equation}
it is easily seen that the minimal modification of the covariant
derivative which satisfies the said properties is obtained by
defining
\begin{eqnarray}
\Omega^{L}_{\mu
\nu}&=&\frac{1}{u^{\beta}n_{\beta}}[a_{\mu}n_{\nu}-a_{\nu}n_{\mu}]\\
&=&[u_{\mu}a_{\nu}-u_{\nu}a_{\mu}]
+\{a_{\mu}[u_{\nu}+\frac{n_{\nu}}{u^{\beta}n_{\beta}}]
-a_{\nu}[u_{\mu}+\frac{n_{\mu}}{u^{\beta}n_{\beta}}]\}.
\end{eqnarray}
The last term represents the angular velocity required to preserve
the direction of light $n$ with respect to the comoving tetrad.

\section{Transformation of photon polarization vectors}
\label{pol}

In this appendix we clarify the relation between this work and
previous works on the transformation properties of the photon
polarization vector. In particular we investigate the relation
between gauge transformations and transformations generated by
$\mathcal{M}$.

The electromagnetic field $F_{\mu \nu}=2\p_{[\mu}A_{\nu]}$ in
vacuum satisfies the Maxwell equation $F^{\mu \nu}_{;\nu}=0$,
that, in terms of the potential $A_{\mu}$, reads
\begin{equation} \label{dym}
 \square A_{\mu}=\p_{\mu}(\p \cdot A).
\end{equation}
By construction this equation is invariant under gauge
transformations $A'_{\mu}=A_{\mu}+\p_{\mu}\alpha(x)$. In this
section we are interested in solutions of the form (plane waves)
\begin{equation}
A_{\mu}(x)=\textrm{Re} [ \epsilon_{\mu} e^{ik_{\nu}x^{\nu}} ]
\end{equation}
where $k^{\mu}$ is a null vector and $\epsilon_{\mu}$ is a complex
vector known as {\em polarization} (for a more general treatment
with a complete spectral decomposition of the general solution and
a momentum dependent polarization see \cite{scaria03}). This plane
wave satisfies Eq. (\ref{dym}) iff $\epsilon^{\nu}k_{\nu}=0$ which
is the equivalent to the Lorentz gauge $\p \cdot A=0$. The Lorentz
gauge is invariant under the restricted gauge transformations such
that $\alpha(x)$ satisfies $\square \alpha=0$. In particular
$\alpha(x)= \textrm{Im}[C e^{ik_{\mu} x^{\mu}}]$, $C \in
\mathbb{C}$, satisfies this condition and hence the transformation
$\epsilon'_{\mu}=\epsilon_{\mu}+C k_{\mu}$ comes from a restricted
gauge transformation. In order to fix the ideas let us take
$k^{\mu}=\omega n^{\mu}=\omega (1,0,\ldots,0,1)$.

 Choose an inertial frame. For
suitable constants $\lambda$, $\tau \in \mathbb{C}$ ($\tau$ is the
analog of $t$), the polarization can be uniquely decomposed in the
form
\begin{equation} \label{poi3}
\epsilon^{\mu}=\begin{pmatrix} \lambda+\tau\\
{\boldsymbol{\epsilon}} \\
\lambda
\end{pmatrix} =
\lambda n^{\mu} + \begin{pmatrix}
\tau\\
\boldsymbol{\epsilon} \\
0
\end{pmatrix}
\end{equation}
where $\boldsymbol{\epsilon}$ is the transverse polarization
vector. We want to find out how the transverse polarization
component $\boldsymbol{\epsilon}$, changes under Lorentz
transformations $G(e_{d+1})$ that preserve the direction of
$n^{\mu}$.

Formally, the problem considered here is very similar to the one
considered previously. However, there are some minor differences:
(i) the group considered does not include the translations, indeed
here we are are considering the subgroup $G(e_{d+1})$ of the
Lorentz group, (ii) despite the fact that there are no
translations the gauge transformations play the role of the
transformations generated by $\mathcal{M}$, in that they add to
$\epsilon^{\mu}$ (the analog of $x^{\mu}$) an arbitrary quantity
proportional to the null vector, (iii) here there is the
additional Lorentz gauge condition $n^{\mu}\epsilon_{\mu}=0$ which
implies $\tau=0$.

Let us come to the solution of the problem. The transformation
$G^{\mu}_{\ \nu}$ has the form given by Eq. (\ref{gma}). Thus, if
$x'^{\mu} = G^{\mu}_{ \ \nu} x^{\nu}$
\begin{equation}
A'^{\mu}=G^{\mu}_{\ \nu} A^{\nu}=\textrm{Re} [{\epsilon'}^{\mu}
e^{ik'_{\beta}x'^{\beta}} ]
\end{equation}
where $k'_{\beta}=e^{-r}k_{\beta}$ and
${\epsilon'}^{\mu}=G^{\mu}_{\ \nu} \epsilon^{\nu}$. From Eqs.
(\ref{gma}) we obtain the transformation of the polarization under
the Lorentz transformation $G^{\mu}_{\ \nu}$ followed by a
restricted gauge transformation of phase $\alpha(x)= \textrm{Im}[C
e^{ik_{\mu} x^{\mu}}]$
\begin{eqnarray}
{\lambda'}&=&\!e^{-r}\lambda+\!\tau (e^{-r}\!\zeta\!\!-\!\sinh
r)-\!e^{-r} \alpha^a \textrm{R}_{ab} \epsilon^{b}+\!C\omega, \\
\tau'&=&\tau e^{r}  \\
{\epsilon'}^{b}&=& R^{b}_{\ c} \epsilon^{c}-\tau\, \alpha^{b},
\qquad b=1,2 \label{sdf}
\end{eqnarray}
We see clearly that if $\lambda=0$ then in general $\lambda'\ne
0$, but $\lambda$  can be sent to zero with a suitable gauge
transformation (choice of $C$). The condition $\tau=0$ implies
$\tau'=0$. In other words the existence of a invariant
simultaneity in Galilean relativity ($t'=e^r t$ implies that the
splitting of spacetime in `simultaneity' slices $t=cnst.$ is
independent of the frame) is related to the Lorentz invariance of
the Lorentz gauge condition ($\tau=0$ $\Rightarrow$ $\tau'=0$). We
conclude that the splitting (\ref{poi3}) is invariant even with
$\tau=\lambda=0$ provided the Lorentz transformation in
$G(e_{d+1})$ is followed by a suitable gauge transformation
\cite{kupersztych76,han81}. The effect of the transformation on
$\boldsymbol{\epsilon}$ amounts to a rotation given by Eq.
(\ref{sdf}) with $\tau=0$. Note that if $\tau=0$ and $R^{a}_{\ b}
=I^{a}_{\ b}$, the parameter $\lambda$ changes while $\tau$ and
$\boldsymbol{\epsilon}$ do not. This fact
 led some authors \cite{han81,kim01,banerjee01,scaria03} to the
 conclusion that the generators $W_a$ generate gauge
transformations instead of Galilean boosts.

In the study of the polarization and its transformation properties
the group $N$ of the previous sections is  replaced  by a subgroup
of the restricted gauge transformations. The invariance of
Galilean simultaneity, $t=cnst$, is replaced by the invariance of
the Lorentz condition and, due to this same  condition ($\tau=0$),
the Galilean group is not fully appreciated since it reduces to
the group of rotations in a  $d$-dimensional Euclidean space.
Indeed, the invariance of $\boldsymbol{\epsilon}$ under Galilean
boosts (\ref{lk2}) was noticed already in \cite{janner71} but, as
we have shown, the role of the Galilei group could not emerge from
studies of the polarization.


\begin{thebibliography}{10}

\bibitem{banerjee01}
Banerjee, R. and Chakraborty, B.: The translation groups as
generators of gauge
  transformation in {$B \wedge F$} theory.
\newblock Phys. Lett. B \textbf{502}, 291--299 (2001)

\bibitem{bargmann54}
Bargmann, V.: On unitary ray representations of continuous groups.
\newblock {A}nn. of {M}ath. \textbf{59}, 1--46 (1954)

\bibitem{bernal03b}
Bernal, A.~N. and {S\'a}nchez, M.: Leibnizian, {G}alilean and
{N}ewtonian
  structures of space–time.
\newblock J. Math. Phys. \textbf{44}, 1129--1149 (2003)

\bibitem{bose95}
Bose, S.~K.: The {G}alilean group in 2 + 1 space-times and its
central
  extension.
\newblock Commun. Math. Phys. \textbf{169}, 385--395 (1995)

\bibitem{bouchiat71}
Bouchiat, C., Fayet, P., and Meyer, P.: Galilean invariance in the
infinite
  momentum frame and the parton model.
\newblock Nucl. Phys. B \textbf{34}, 157--176 (1971)

\bibitem{dirac49}
Dirac, P. A.~M.: Forms of relativistic dynamics.
\newblock Rev. Mod. Phys. \textbf{21}, 392--399 (1949)

\bibitem{duval85}
Duval, C., Burdet, G., K{\"u}nzle, H.~P., and Perrin, M.: Bargmann
structures
  and {N}ewton-{C}artan theory.
\newblock Phys. Rev. D \textbf{31}, 1841--1853 (1985)

\bibitem{duval91}
Duval, C., Gibbons, G., and Horv\'athy, P.: Celestial mechanics,
conformal
  structures, and gravitational waves.
\newblock Phys. Rev. D \textbf{31}, 1841--1853 (1985)

\bibitem{ehlers97}
Ehlers, J.: Examples of {N}ewtonian limits of relativistic
spacetimes.
\newblock Class. Quantum Grav. \textbf{14}, A119–--A126 (1997)

\bibitem{eisenhart29}
Eisenhart, L.~P.: Dynamical trajectories and geodesics.
\newblock Ann. Math. (Ser 2) \textbf{30}, 591--606 (1929)

\bibitem{elizade78}
Elizade, E. and Gomis, J.: The groups of {P}oincar\'e and {G}alilei
in
  arbitrary dimensional spaces.
\newblock J. Math. Phys. \textbf{19}, 1790--1792 (1978)

\bibitem{han81}
Han, D. and Kim, Y.~S.: Little group for photons and gauge
transformations.
\newblock Am. J. Phys. \textbf{49}, 348--351 (1981)

\bibitem{harindranath96}
Harindranath, A.: \emph{An Introduction to Light-Front Dynamics for
  Pedestrians}, Ames: International Institute of Theoretical and Applied
  Physics, vol. Light-Front Quantization and Non-Perturbative QCD (1996)

\bibitem{janner71}
Janner, A. and Janssen, T.: Electromagnetic compensating gauge
transformations.
\newblock Physica A \textbf{53}, 1--27 (1971)

\bibitem{julia95}
Julia, B. and Nicolai, H.: Null-{K}illing vector dimensional
reduction and
  {G}alilean geometrodynamics.
\newblock Nucl. Phys. B \textbf{439}, 291--323 (1994)

\bibitem{kim01}
Kim, Y.~S.: Internal space-time symmetries of massive and massless
particles
  and their unification.
\newblock Nucl. Phys. B \textbf{102/103}, 369--376 (2001)

\bibitem{kunzle72}
K{\"u}nzle, H.~P.: Galilei and {L}orentz structures on space-time:
comparison
  of the correspondig geometry and physics.
\newblock Ann. Inst. H. {P}oincar\'e Phys. Theor. \textbf{17}, 337--362 (1972)

\bibitem{kupersztych76}
Kupersztych, J.: Is there a link between gauge invariance,
relativistic
  invariance and electron spin?
\newblock Il {N}uovo {C}imento \textbf{31}, 1--11 (1976)

\bibitem{kutach04}
Kutach, D.: Non-relativistic quantum mechanics on a {K}aluza-{K}lein
manifold
  (2004)

\bibitem{leutwyler78}
Leutwyler, H. and Stern, F.: Relativistic dynamics on a null plane.
\newblock Ann. Phys. \textbf{112}, 94--164 (1978)

\bibitem{levi96}
Levi, M.: Composition of rotations and parallel transport.
\newblock Nonlinearity \textbf{9}, 413--419 (1996)

\bibitem{levyleblond63}
L\'evy-Leblond, J.~M.: Galilei group and nonrelativistic quantum
mechanics.
\newblock J. Math. Phys. \textbf{4}, 776--788 (1963)

\bibitem{lichnerowicz55}
Lichnerowicz, A.: \emph{Th\'eories relativistes de la gravitation et
de
  l'{\'e}lectromagnetisme, {R}elativit{\'e} {G\'en\'erale} et th\'eories
  unitaires}.
\newblock Paris: Masson (1955)

\bibitem{marsden99}
Marsden, J.~E. and Ratiu, T.~S.: \emph{Introduction to Mechanics and
Symmetry}.
\newblock New York: Springer (1999)

\bibitem{minguzzi04c}
Minguzzi, E.: Differential aging from acceleration: {A}n explicit
formula.
\newblock Am. J. Phys. \textbf{73}, 876--880 (2005)

\bibitem{naber92}
Naber, G.~L.: \emph{The geometry of {M}inkowski spacetime}.
\newblock New York: {Springer-Verlag} (1992)

\bibitem{oraifeartaigh00}
{O'R}aifeartaigh, L. and Straumann, N.: Gauge theory: Historical
origins and
  some modern developments.
\newblock Rev. Mod. Phys. \textbf{72}, 1--23 (2000)

\bibitem{penrose59}
Penrose, R.: The apparent shape of a relativistically moving sphere.
\newblock Proc. Camb. Phil. Soc. \textbf{55}, 137--139 (1959)

\bibitem{rindler84}
Penrose, R. and Rindler, W.: \emph{Spinors and space-time. {V}ol.
1}.
\newblock Cambridge Monographs on Mathematical Physics. Cambridge: Cambridge
  University Press (1984)

\bibitem{scaria03}
Scaria, T.: Translational groups as generators of gauge
transformations.
\newblock Phys. Rev. D \textbf{68}, 105013 (2003)

\bibitem{schiller03}
Schiller, C.: Motion mountain - the physics textbook (2003).
\newblock Available at www.motionmountain.net

\bibitem{weinberg64}
Weinberg, S.: Feynman rules for any spin. {II}. {M}assless
particles.
\newblock Phys. Rep. \textbf{134}, B882--B896 (1964)

\bibitem{weinberg95}
Weinberg, S.: \emph{The Quantum Theory of Fields}, vol.~I.
\newblock Cambridge: Cambridge {U}niversity {P}ress (1995)

\end{thebibliography}

\end{document}